\documentclass[bibyear]{aa} 

\graphicspath{{img/}}
\usepackage{graphicx}
\usepackage{caption}
\usepackage{subcaption}
\usepackage{float}
\usepackage{comment}
\usepackage{natbib}
\usepackage{txfonts}
\usepackage{pdflscape}
\usepackage{longtable}
\usepackage{rotating}
\usepackage{amsmath}
\usepackage{tabularx}
\usepackage{multirow}
\usepackage{xcolor}
\usepackage{rotating}
\usepackage{booktabs}
\usepackage[symbol]{footmisc}

\usepackage{hyperref}
\usepackage{hyperref}
\usepackage{etex}
\hypersetup{%
    colorlinks=true, 
    citecolor=blue,
    filecolor=blue,
    linkcolor=blue,
    urlcolor=blue
    }

\begin{document} 

   \title{Multiwavelength properties of short gamma ray bursts with extended emission observed by \textit{Swift}}

   \author{
          M.M.~Dinatolo\inst{1,2}
          \and
          A.~Mei\inst{2}
          \and
          R.~Brivio\inst{2}
          \and
          M.~Ferro\inst{2}
          \and
          M.G.~Bernardini\inst{2}
          \and
          P.~D'Avanzo\inst{2}
          \and
          S.~Belova\inst{3}
          \and
          S.~Campana\inst{2}
          \and
          S.~Covino\inst{2}
          \and
          D.~Frederiks\inst{3}
          \and
          B.~Haskell\inst{1,4}
          \and
          R.~Salvaterra\inst{2}
          \and
          B.~Sbarufatti\inst{2}
          \and
          A.~Tsvetkova\inst{3}
}

    \institute{
            Università degli Studi di Milano, Dipartimento di Fisica, Via Celoria 16, 20133 Milan, Italy 
        \and
             INAF – Osservatorio Astronomico di Brera, Via E. Bianchi 46, 23807 Merate (LC), Italy
        \and
            Ioffe Institute, Politekhnicheskaya 26, 194021, St. Petersburg, Russia
        \and
            INFN Sezione di Milano, Via Celoria 16, 20133 Milan, Italy 
            }
            
    \date{Received xxx; accepted yyy}


\abstract
  {
  Short gamma-ray bursts with extended emission (SGRBEEs) are a particular class of long GRBs (LGRBs) which, despite their duration, share several observational features with short GRBs (SGRBs). They are composed by a short, hard initial pulse (IP) followed by a longer and softer extended emission (EE). These events challenge the traditional duration-based classification, which usually links LGRBs to collapsars and SGRBs to neutron star mergers.} 
  {We investigate whether SGRBEEs originate from the same progenitor as SGRBs despite their duration, representing a peculiar subclass of LGRBs, or if they constitute a distinct population. Given their respective duration, we tested if the IP and the EE share properties with short and long GRBs, respectively.} 
  {We analysed SGRBEEs from the flux-limited, redshift-complete SBAT$4$ sample using prompt-emission data from \textit{Swift}/BAT, \textit{Fermi}/GBM and Konus-WIND, and X-ray afterglow observations from \textit{Swift}/XRT. The temporal and spectral properties of the IP and EE components were compared with those of SGRBs from the extended SBAT4 sample and LGRBs from the BAT6 sample. } 
  {Despite observing a clear spectral evolution during the prompt phase of each SGRBEE, IPs and EEs as well as short and long GRBs can not be distinguished by their hardness ratio only. 
  All bursts analysed have a spectral lag consistent with zero. SGRBEEs X-ray light curves are consistently more complex than those of short bursts, requiring the addition of multiple breaks and showing the presence of steep decays and plateaus.} 
  {Our results indicate that SGRBEEs are not an intermediate class. During prompt emission, despite common spectral features, IPs and EEs temporally differ from short and long GRBs, respectively. EEs are on average too faint to appear in the Amati plane, but IPs occupy the same parameter space of SGRBs. 
  In the afterglow, SGRBEEs are more luminous than standard GRBs at early-time, suggesting a direct contribution from the EE; at later times, these events behave as standard SGRBs, and both remain systematically less luminous than LGRBs.}
      
   
    

   \keywords{gamma rays bursts - high energy astrophysics - astroparticle physics
               }


   \maketitle

%
\section{Introduction} \label{intro}
Gamma ray bursts (GRBs) are rapid flashes of gamma rays occurring at cosmological distances. Their emission is divided into two phases: \textit{prompt} and \textit{afterglow}. During the initial prompt phase, the GRB releases most of its energy, mainly in the gamma-ray band \citep{piran_04}. It lasts from milliseconds to several minutes, and it is characterized by fast temporal variability \citep{Bhat_2013}. The emission spectrum is broad and non-thermal, typically modeled phenomenologically using power law functions as it originates from accelerated particles whose energy distribution follows a power law, as expected from shock acceleration processes \citep{band93}. The prompt emission is commonly interpreted as originating from internal dissipation processes within an ultra-relativistic jet launched by a compact central engine \citep{Sari1999}, possibly through internal shocks \citep{Rees_1994}, magnetic reconnection \citep{Drenkhahn_2002, Zhang_2011a}, or sub-photospheric processes \citep{Rees_2005}. The afterglow can be observed from the radio band up to GeV energies and typically shows a smooth decay over timescales of days to weeks \citep{Zhang_2007}. It is generally interpreted as synchrotron radiation produced when the relativistic outflow interacts with the surrounding circum-burst medium, generating an external shock \citep{Meszaros_1997}. GRBs are traditionally classified into two main categories based on the duration of their prompt emission ($T_{\text{90}}$, \citealt{Kouveliotou93}): long GRBs (LGRBs; $T_{\text{90}} \gtrsim 2$ s), typically associated with the collapse of massive stars \citep{Woosley93, Woosley06, MacFadyen_1999}, and short GRBs (SGRBs; $T_{\text{90}} \lesssim 2$ s), linked to compact binary mergers \citep{Eichler89, Narayan92, Berger14, Abbott17, Goldstein17, Cowperthwaite17, Abbott17_2, Savchenko17, evans17}.
\par However, the growing number of observed events has revealed a population that challenges this simple bimodal distribution, such as a SGRB associated with a supernova (GRB\,200826A, \citealt{Rossi_22}), and long-duration GRBs with related kilonova emission (GRB\,211211A, \citealt{Rastinejad_22, Troja_22}, and GRB\,230307A, \citealt{Levan_24}). An additional peculiar case is represented by SGRBs with extended emission (SGRBEEs), which show an initial short and hard spike followed by a long lasting and softer tail, whose physical origin is still debated \citep{Norris_2006, Gehrels_2006, Norris11, Metzger_2008, Gompertz_2013}. Extended emission (EE) was first systematically identified in BATSE data by \citet{Gehrels_2006}, who showed that a significant fraction of SGRBs display a temporally distinct soft component lasting up to hundreds of seconds. Subsequent studies confirmed that this emission is spectrally softer than the initial pulse (IP) and can even dominate, in terms of fluence, the prompt emission \citep{Kaneko_2015}. To explain its unclear physical origin several scenarios have been proposed, including prolonged central engine activity powered by fall-back accretion \citep{Metzger_2008}, the formation of a long lived magnetar remnant \citep{Gompertz_2013}, or late time internal dissipation within the relativistic outflow. Determining the correct burst duration is therefore crucial when inferring the progenitor of these events. If the EE is included in the $T_{\text{90}}$ measurement, many SGRBEEs would formally fall into the long-duration category despite showing hints of compact-object mergers. This highlights a key limitation of duration-based classifications and motivates a more detailed investigation of the temporal and spectral structure of GRBs.

For the prompt phase, time-resolved spectral analysis has shown a clear hard-to-soft evolution between the IP and the EE (\citealt{Kaneko_2015}), while their position in the hardness–duration plane often places the two phases in regions typically occupied by short and long GRBs, respectively \citep{Perley_09}. The afterglow properties provide an additional key diagnostic. X-ray studies have shown that SGRBEEs can display plateau phases and complex light curve morphologies, possibly indicating sustained energy injection, as expected in magnetar central-engine scenarios \citep{Gompertz_2013}. However, systematic comparisons with the general SGRB population are often limited by selection effects and by the lack of samples with high redshift completeness \citep{Berger14}. 

A step forward in the study of GRB populations has been the construction of complete, flux-limited samples with high redshift completeness, such as BAT$6$ for LGRBs \citep{Salvaterra12} and SBAT$4$ for SGRBs \citep{Davanzo14}, which provides the highest redshift completeness currently available for SGRBs. These samples were selected from the population of GRBs detected by the Neil Gehrels Swift Observatory (\textit{Swift}, \citealt{Gehrels_04}). They minimize observational biases and allow statistically meaningful comparison between different GRBs. In this work, we focus on SGRBEEs in order to assess whether they form a distinct population, a link class, or simply the tail of the SGRB distribution. Despite the numerous observations, a comprehensive study that investigates the prompt spectral evolution, the energetics, and the X-ray afterglow properties of SGRBEEs within a complete sample is still missing. 
We recently extended the orignal SBAT4 sample up to Jan. 2026, including all SGRBs satisfying our selection criteria (see Sect.\ref{sec:sample}) and reaching a size that is comparable to the BAT6 sample for LGRBs. The extended sample is presented in a companion paper (D'Avanzo et al., in preparation), where the details of the sample definition and extension and the analysis of the prompt phase properties are presented. Two additional companion works focus on a complete analysis of the afterglow and host galaxy properties (Brivio et al., Ferro et al., in preparation, respectively).
\par Here we present a systematic multi-wavelength characterization of SGRBEEs selected from the extended SBAT4 sample. We perform a time-resolved analysis of the prompt emission observed by the Burst Alert Telescope (BAT; $15-150$ keV, \citealt{Barthelmy_05}) on board the \textit{Swift} satellite, separating the IP from the EE and comparing their spectral and temporal properties with those of the SBAT$4$ extended SGRBs and BAT$6$ LGRBs populations. We then study the spectral lag. For the events also detected by the
Gamma-ray Burst Monitor (GBM; $8\ \text{keV} - 40$ MeV, \citealt{Meegan_09}) on board the \textit{Fermi} satellite and the Konus gamma-ray spectrometer on board the Wind spacecraft (KW; $\sim20\ \text{keV} - 20\ \text{MeV}$, \citealt{Aptekar1995}), we derive the broadband energetics and investigate their position in the $E_{\text{peak, z}} - E_{\text{iso}}$ plane, the so-called \textit{Amati} plane. $E_{\text{peak, z}}$ is the peak of the spectrum in the spectral energy distribution (SED) representation in rest frame, and $E_{\text{iso}}$ is the isotropic equivalent energy \citep{Amati_2002, amati06, Amati_2008}. We study the X-ray afterglow light curves in both the observer and rest frames, including a morphological classification and a statistical comparison with the complete short and long GRB samples.

\section{Methods}\label{sec:methods}
\subsection{Sample selection}\label{sec:sample}
We analysed GRBs observed by the \textit{Swift} satellite that are part of the extended SBAT4 sample (D'Avanzo et al., in preparation). The sample includes events with particularly favorable observational conditions (including photometric redshfit estimates): GRBs with  $T_{\text{90}} < 2$ s as well as \textit{Swift} GRBs with $T_{\text{90}} > 2$ whose BAT light curve shows a short-duration peak followed by a softer, long-lasting tail (i.e., the extended emission), according to the prescriptions presented in \cite{Lien_17}. After this initial selection, the extended SBAT4 sample includes only those events which:
\begin{itemize}
    \item[i.] are re-pointed by the \textit{Swift} X-Ray telescope (XRT; $0.3 - 10$ keV, \citealt{Burrows_05}) within $120$ s from the BAT trigger;
    \item[ii.] have low Galactic extinction in the direction of the burst ($A_V < 1.2$ mag);
    \item[iii.] show a $1$s peak photon flux $P \ge 3.5$ photons s$^{-1}$ cm$^{-2}$, corresponding to an instrument that is $\sim 4$ times less sensitive than BAT, and computed using $15 - 150$ keV BAT light curves binned with $\delta t = 64$ ms.
\end{itemize} 

The sample consists of $53$ SGRBs, $79\%$ of which have a measured redshift. Of these, $11$ GRBs have a $T_{\text{90}}>2$ s, which we consider our sample of interest for SGRBEEs. These events therefore represent approximately $20\%$ of the SGRBs in the extended SBAT4 sample. Although statistically consistent, considering the entire sample of SGRBs is not representative of the short population. The description of their physical properties requires the measurement of the redshift, which is available only for about $50\%$ of all events. However, the extended SBAT4 sample allows us to achieve the highest possible redshift completeness of $79\%$. 
Furthermore, we adopt the BAT6 sample \citep{Salvaterra12, Nava_2012} as the reference population for LGRBs. This sample comprises $58$ bursts observed up to May 2011, achieving a redshift completeness level of $\sim 90\%$. The extended SBAT4 sample is extracted using identical selection criteria, with the sole exception that the $1$ s peak photon flux threshold for LGRBs is set to $P = 2.6$ photons s$^{-1}$ cm$^{-2}$. The comparable sample sizes and redshift completeness levels of the BAT6 and extended SBAT4 samples facilitate a direct comparison (Table~\ref{tab:sample_pres}), which we present in Sect.~\ref{sec:results}.
\begin{table}[tbh]
\centering
\caption{Composition of the samples.}
\label{tab:sample_pres}
\begin{tabular}{lcc}
\toprule\toprule

Sample & Total & With redshift \\[0.5ex]

\midrule

SBAT$4$ (all events)      & 53 & 42 (79\%) \\[0.5ex]
\quad without EE          & 42 & 32 (76\%) \\[0.5ex]
\quad with EE (SGRBEE)    & 11 & 10 (91\%) \\[0.5ex]

\cmidrule(lr){1-3}

BAT6                      & 58 & 52 (90\%) \\[0.5ex]

\bottomrule
\end{tabular}
\end{table}

\subsection{Data reduction}\label{sec:datared}
\subsubsection{\textit{Swift}} \label{sec:swift}
The data collected by the \textit{Swift} satellite are publicly available on the mission's official website or through the High Energy Astrophysics Science Archive Research Center (HEASARC) portal\footnote{\url{https://www.swift.ac.uk/swift_portal/}}. The software used for data analysis is HEASoft\footnote{\url{https://heasarc.gsfc.nasa.gov/docs/software/lheasoft/}} (version $6.9$), along with the Calibration Database (CALDB) version $20180925$. Spectral analysis was performed using the XSPEC software\footnote{\url{https://heasarc.gsfc.nasa.gov/docs/software/xspec/manual/index.html}}, version $12.12.0$.
BAT data were downloaded from the \textit{Swift} data archive. The FTOOL pipeline \texttt{batgrbproduct} is used to extract scientific products such as light curves, images, and spectra on standard time intervals and energy bands. Using the FTOOLS pipelines \texttt{batmaskwtevt} and \texttt{batbinevt}, we extracted the background-subtracted mask-weighted BAT light curves. We used the FTOOLS pipelines \texttt{batbinevt}, \texttt{batupdatephakw}, \texttt{batphasyserr}, and \texttt{batdrmgen} to produce BAT standard spectra and corresponding response files.
\par We downloaded the XRT data from the \textit{Swift} Burst Analyser Website\footnote{\url{https://www.swift.ac.uk/burst_analyser/}} \citep{Evans_10}. We considered the observation both in the window timing (WT) and photon counting (PC) modes, retrieving the observed unabsorbed flux light curves in the $0.3-10$ keV energy band. 

\subsubsection{\textit{Fermi}}\label{sec:fermi}
 \textit{Fermi}/GBM data are extracted from the application programming interface \textit{Fermi} GBM Data Tools\footnote{\url{https://fermi.gsfc.nasa.gov/ssc/data/analysis/gbm/gbm_data_tools/gdt-docs/index.html}}.
For each spectrum, we selected the two sodium iodide detectors (NaI, $8-900$ keV) and one bismuth germanate detector (BGO, $300$ keV$- 40$ MeV) with lowest separation angle from the source. We estimated the background through a polynomial fit, selecting time intervals far from the event emission. We obtained the spectrum, the response matrix, and the background integrated over the same time interval identified for the BAT data.

\subsubsection{Konus-WIND}\label{sec:kw}
Konus-Wind is a gamma-ray spectrometer consisting of two identical NaI(Tl) detectors, S1 and S2, which observe the southern and northern ecliptic hemispheres, respectively. Only one detector can be triggered at a time providing high-resolution spectral data collected in two overlapping energy intervals, PHA1 ($\sim$20–1300~keV) and PHA2 ($\sim$250~keV–16~MeV). The total duration of the spectral measurements is up to 490 s. The KW background is very stable and assumed to be at a constant level during the triggered mode record. The light curves are available, starting from T0(KW)-0.512~s, in three energy windows G1 ($\sim$20–80~keV), G2 ($\sim$80–300~keV), and G3 ($\sim$300–1300 keV), with time resolution varying from 2 up to 256~ms and a total record duration of $\sim$230~s. The spectral interval start and end times were corrected for the delay time corresponding to the propagation of signal from the Wind spacecraft (at L1) to the \textit{Swift} spacecraft; if the value is positive, the signal first arrives at Wind. The light curves were shifted correspondingly. More information on the KW data formats and data reduction procedure may be found in \citet{Tsvetkova2017, Tsvetkova2021}, the KW sample of SGRBs (including SGRBEEs) can be found in \citet{Svinkin2016, Lysenko2025}.

\subsection{Spectral fit routine} \label{sec:fit_routine}
After the data reduction, we performed time-resolved spectral analysis of the GRBs in our sample using BAT, GBM and KW data. For the fitting process, we used the Heasarc package \textsc{XSPEC} \citep{Arnaud_96}.
\par We ignored the energy channels outside $15-150$ keV for the BAT detector: beyond this range, the detector becomes transparent to radiation. For the GBM data, we ignored the energy channels outside the $8-900$ keV range for the NaI detectors, as well as the $30-40$ keV band in order to avoid the iodine K-edge line at 33.17 keV \citep{Meegan_2009}. We ignored the energy channels outside $300\ {\rm keV}- 40$ MeV for the BGO detectors. For the Konus-WIND data, we removed the overlapping channels of the PHA1 and PHA2 data. 

Initially, to analyse the prompt spectra of GRBs in our sample, we included only BAT spectra in our dataset. For the estimation of the $E_{\text{peak,z}} - E_{\text{iso}}$ (Amati) correlation, we considered all the data available. In this case, during the spectral fit routine, we included the presence of a cross-calibration constant, \texttt{constant} in XSPEC notation, allowing for a $30\%$ variation for each dataset. We applied Gaussian statistics (\texttt{chi}) to BAT data, Poisson-Gaussian statistics (\texttt{pgstat}) to GBM data and Cash statistics (\texttt{cstat}) to KW data. We extracted all synchronous spectra using the \textit{Swift} trigger time as a reference. Time intervals adopted for the two phases differ from those used in the spectral analysis based solely on BAT data, owing to the spectral extraction methodology applied to the KW observations. We therefore re-extracted the \textit{Swift} spectra using these revised temporal intervals.
\par For each event, we identified two distinct phases: an IP and an EE. In order to distinguish between these two, we binned each light curve according to their signal-to-noise ratio (SNR) setting a threshold of $ \text{SNR}= 3$. We consider the IP and EE intervals as contiguous, and the separation time, $ T_{\rm cut}$ was derived from a visual check of SNR light curves. The initial time of the IP segment coincides with BAT trigger time. Conversely, the EE segment ends at the end of the $T_{\text{90}}$ interval. We extracted a spectrum during each of the two phases, between  $[T_{90, \rm start}, T_{90, \rm start} + T_{\rm cut}]$ and $[T_{90, \rm start} + T_{\rm cut}, T_{90, \rm start} + T_{90}]$, respectively.
\par GRB prompt emission spectra are broad and characterized by peak in their spectral energy distribution ($\nu F_{\nu}$). However, the choice of the phenomenological model used to describe these spectra depends on the energy band of the observing instruments. When a GRB is observed by BAT only, which operates in a relatively narrow and soft energy band, the spectral peak often falls outside its sensitivity range. In these cases, BAT photon spectra, $\rm N(E)$, are adequately described by a simple power law model (PL). In some cases, the spectral peak might enter the BAT energy band, but the instrument's limited high-energy coverage prevents the detection of the second power-law component. To properly estimate the peak in these scenarios, we introduced an exponential damping, using the cutoff power law model (CPL). The goodness of fit is evaluated using the $\chi^2$ statistics. We used the Akaike Information Criterion (AIC; \citealt{akaike_74}) to determine whether the introduction of an exponential cutoff statistically improves the fit. The AIC value of the model considered is defined as: 
\begin{equation}
    AIC = 2k - 2 \ln \mathcal{L} = 2k + \chi^2 ,
\end{equation}
where $k$ is the number of estimated parameters in the model and $\mathcal{L}$ is the likelihood of the model. If $\Delta AIC = AIC_{PL}  - AIC_{CPL} < 2$  we considered the power law as the best fit model, otherwise we consider the presence of the exponential cutoff to better represent the data.
\par For those events jointly observed by GBM and KW, the broader energy coverage allowed us to observe the full spectral shape, including the power law below the peak and the high energy tail. This enabled the use of more complex spectral models, such as the Band function \citep{band93}.
\par In addition, we computed for each GRB the isotropic-equivalent energy, $E_{\text{iso}}$, as:
\begin{equation}
    E_{\text{iso}} = 4 \pi d_L^2 \cdot \frac{T_{\text{90}}}{1+z} \cdot \int_{a}^{b} N(E) \cdot E \ dE , 
\end{equation}
where $\rm N(E)$ is the spectral model, and $d_L(z)$ is the luminosity distance at the redshift of the event, calculated using the cosmological parameters of \cite{planck_18}. We used an integration interval between $a=1$ keV and $b=10$ MeV in the source frame. To compute them in the rest frame, we divided $a$ and $b$ by a factor $(1+z)$.
\par In order to robustly explore the parameter space and estimate parameter uncertainties, we ran a Markov Chain Monte Carlo (MCMC) using the XSPEC command \texttt{chain}. Chains were used to sample the posterior distribution of the model parameters. The best–fit values reported in this work correspond to the median of the posterior distributions obtained from the MCMC. Parameter uncertainties are derived from the 16th and 84th percentiles of the posterior distributions, corresponding to the $68\%$ credible interval. Derived quantities such as $E_{\text{peak,z}}$ and $E_{\text{iso}}$ were computed for each chain step and their uncertainties were estimated from the resulting distributions. Moreover, we computed the hardness ratio, $HR$,  during both the IP and EE phases, as the ratio of fluxes, $F$, in the hard ($50-100$ keV) and soft ($25-50$ keV) energy bands:
\begin{equation}
    HR = \frac{F_{50-100}}{F_{25-50}}\ .
\label{eq:hr}
\end{equation}

We report the results of the BAT time-resolved spectral analysis of the SGRBEEs in our sample, during both the IP and EE phases, in Table~\ref{tab:results}. To make a comparison with the GRBs present in the SBAT$4$ and BAT$6$ samples, we performed the same analysis for all these events. The spectra were extracted considering the entire duration, estimated through the  $T_{\text{90}}$.

\subsection{Spectral lag}\label{sec:lag}
Spectral lag is defined as the time delay between the arrival of photons in different energy bands within a single GRB pulse \citep{cheng_95}. By convention, the lag is positive when high energy photons precede low energy ones. The primary explanation for this phenomenon is the pulse hard-to-soft evolution \citep{Dermer_1998, Ryde_2005}. As the pulse evolves, the spectral peak energy decreases over time, shifting to lower energy bands \citep{Ford_1995}. Alternatively, it is explained as a curvature effect \citep{Salmonson_2000, Lu_2006}: photons emitted at larger angles relative to the observer's line of sight must travel a slightly longer path, arriving with a delay. Due to the relativistic Doppler effect, this delayed, off-axis emission is also redshifted to lower energies, producing a positive spectral lag \citep{Salmonson_2000}. \par The spectral lag has been used as a possible tool to discriminate between LGRBs and SGRBs. LGRBs typically exhibit significant, measurable positive lags \citep{Gehrels_2006}. Also, they follow the lag-luminosity relation, an anti-correlation demonstrating that bursts with higher peak isotropic luminosity have shorter spectral lags \citep{Norris_2000}. In contrast, SGRBs exhibit very small lags that are consistent with zero, possibly indicating rapid dissipation processes typical of compact object mergers \citep{Norris_2006}. However, \cite{Bernardini_2015} analysed 50 bright LGRBs and 6 SGRBs observed by BAT and found that half of the long ones have a lag consistent with zero. Also, SGRBs overlap with the zero-lag long GRB population in the lag-luminosity plane. 
\par Following \cite{Bernardini_2015}, for each GRB in our sample (Table~\ref{tab:results}), we extracted mask-weighted, background subtracted light curves 
in two energy bands corresponding to the rest frame intervals 25-50 keV (hereafter ch1) and 50-100 keV (hereafter ch2). We have analysed each GRB light curve using data from both ch1 and ch2. For each energy channel, we considered two time intervals, corresponding respectively to the IP and EE.
\par In order to measure the temporal correlation of the two light curves in ch1 and ch2, we used a discrete cross correlation function (CCF), adopting its non-mean subtracted formula, as reported in  \cite{Band_1997}:
\begin{equation}
    \text{CCF}(k\Delta t; c_1, c_2) = \frac{\sum_{i = max(1, 1-k)}^{min(N, N-k)} c_{2i}\ c_{1(i+k)}}{\sqrt{\left(\sum_i\ c_{1i}^2\right) \cdot  \left( \sum_i\  c_{2i}^2\right)}}, 
\end{equation}
where $\Delta t$ is the duration of the time bin, $k\Delta t$ ($k = -N$..., $-1$, $0$, $1$, ..., $N$) is a multiple of the time bin and represents the time delay, $N$ is the total number of data point considered in the light curves, $c_1$ and $c_2$ are the count rates of ch1 and ch2, respectively.

For each light curve in each channel, we calculated the CCF value for a series of time delays and we defined the spectral lag, $\tau$, as the time delay corresponding to the global maximum of the CCF
. To obtain this value, we fitted an asymmetric Gaussian model to the CCF:
\begin{equation}
    \text{CCF}(k\Delta t) = A \cdot \left \{ \begin{array}{rl}  exp \left[-\frac{(k\Delta t-\tau)^2}{2 \Sigma_l^2}\right]  & k\Delta t \le \tau       \\
     exp \left[-\frac{(k\Delta t-\tau)^2}{2 \Sigma_r^2}\right]  & k\Delta t > \tau 
\end{array}
\right.     ,
\end{equation}

where $A$ is the normalization, and $\Sigma_l^2$ and $\Sigma_r^2$ are the left and right uncertainties, respectively. 
The uncertainties on the CCF have been derived by applying a flux-randomization method. We generated $10 000$ realizations of ch1 and ch2 light curves based on each count rate $\bar c_i$ and its error $\Delta c_i$:
\begin{equation}
    c_i = \bar c_i + \xi _i \Delta c_i,
\end{equation}
where $\xi _i$ is a random number drawn from a standard normal distribution. For each time delay $k\Delta t$, the corresponding value of the CCF and its error are the mean and the standard deviation of the distribution of the CCF calculated from the $10000$ realizations of the light curves. We derived the uncertainty on the spectral lag from the fit to 1000 realizations of the CCF versus time delay with the same randomization method. We started using mask-weighted light curves binned at $\Delta t = 2$ ms and doubled the bin size up to $\Delta t = 1.024$ s. The optimal bin size was selected as the smallest bin for which both energy channels had a sufficient SNR. We also checked whether $\text{CCF}_{max} \ge 0.5$, ensuring that the correlation is strong enough and not dominated by statistical noise, thus providing a reliable measurement of the spectral lag \citep{Ukwatta_2010}.

\subsection{XRT observations of the X-ray afterglow} \label{sec:xrt}
Most of GRBs X-ray light curves consist of power laws segments with zero or multiple breaks. To study the XRT light curves of GRBs in our sample, we used the same models presented by \cite{margutti2012}, defined as: 
\begin{equation}
    f(N_1, \gamma_1, t) = N_1 \cdot t ^{- \gamma_1} , 
\end{equation}
\begin{equation}
    g(N_2, \gamma_2, \gamma_3, t_b, s, t) = N_2 \cdot \left(\left(\frac{t}{t_b}\right) ^{- \frac{\gamma_2}{s}} + \left(\frac{t}{t_b}\right) ^{- \frac{\gamma_3}{s}}\right)^s .
\end{equation}

Both $f$ and $g$ represent the X-ray flux. The function $f$ describes a light curve with a single power law segment with normalization $N_1$ and slope $\gamma_1$. Instead, the function $g$ describes a light curve with normalization $N_2$ constituted by two smoothed power law segments. $\gamma_2$ and $\gamma_3$ are the temporal indices before and after the break time $t_b$, respectively, while $s$ is the smoothness parameter. We fitted a combination of both the \textit{f} and \textit{g} functions to describe X-ray light curves, namely:

\begin{itemize}
    \item \textbf{Model 0}, simple power law:
    \begin{equation}
        F = f(N_1, \gamma_1, t) , 
    \end{equation}
    \item \textbf{Model I}, smoothed broken power law:
    \begin{equation}
        F = g(N_1, \gamma_1, \gamma_2, t_{b1}, s_1, t) , 
    \end{equation}
    \item \textbf{Model IIa}, smoothed broken power-law plus initial power law decay:
    \begin{equation}\label{eq:XLC_modelIIa}
        F = f(N_1, \gamma_1, t) + g(N_2, \gamma_2, \gamma_3, t_{b2}, s_1, t), 
    \end{equation}
    \item \textbf{Model IIb}, smoothed broken power-law plus final power law decay:
    \begin{equation}\label{eq:XLC_modelIIb}
        F = g(N_1, \gamma_1, \gamma_2, t_{b1}, s_1, t) + f(N_2, \gamma_3, t), 
    \end{equation}
    \item \textbf{Model III}, double smoothly joined broken power laws:
    \begin{equation}
        F = g(N_1, \gamma_1, \gamma_2, t_{b1}, s_1, t) + g(N_2, \gamma_3, \gamma_4, t_{b2}, s_2, t), 
    \end{equation}
\end{itemize}
In the nomenclatures of models 0, I, IIa, IIb and III the subscript of the temporal index refers to the order of the power law segment.

For each GRB, we fitted the flux in the observer frame energy band, $0.3-10$ keV, and in the common rest frame energy band, $2-10$ keV, computed from the observed $0.3-10$ keV unabsorbed fluxes $f_x$ and the time resolved photon index $\Gamma$ (retrieved from the \textit{Swift} Burst Analyser) as:
\begin{equation}
    f_x^{crf}(2-10\ {\rm keV}) =  f_X(0.3 - 10\ {\rm keV})\ \frac{\left(\frac{10}{1+z}\right)^{2-\Gamma} - \left(\frac{2}{1+z}\right)^{2-\Gamma}}{10^{2-\Gamma} - 0.3^{2-\Gamma}}\ .
\label{eq:common_frame}
\end{equation}
The parameter estimation was performed using a Markov Chain Monte Carlo (MCMC) approach. We adopted a standard Gaussian log-likelihood function, assuming normally distributed errors on the flux data. We applied bounded priors on the parameters. In particular, the temporal break parameters were strictly constrained to fall within the observational time window of each GRB ($t_{min} < t_b < t_{max}$). The best fit parameters were derived from the posterior distributions, taking the median as the optimal value. The $1 \sigma$ statistical uncertainties were defined by the $16$th and $84$th percentiles of the distributions. Because our model set includes both nested and non nested models, we implemented a model selection algorithm to determine the best fit. For models that are nested we applied the Fisher's test (F-test) to evaluate whether the addition of free parameters significantly improved the fit. The more complex model was preferred only if the resulting p-value was below a threshold of $0.05$. For non-nested models we used the AIC. The model minimizing the information criterion was selected as the global best fit. In cases where the criteria of two models were statistically indistinguishable ($\Delta AIC \le 2$), we adopted the model with fewer free parameters. We identified flux excesses following the flare identification procedure reported in the \textit{Swift} documentation\footnote{\url{https://www.swift.ac.uk/xrt_live_cat/docs.php}}. We applied the same procedure for the SGRBEEs and for the extended SBAT4 GRBs.

\section{Results}\label{sec:results}
\subsection{Prompt phase} \label{sec:prompt_res}
\begin{figure}
    \centering
    \includegraphics[width=\linewidth]{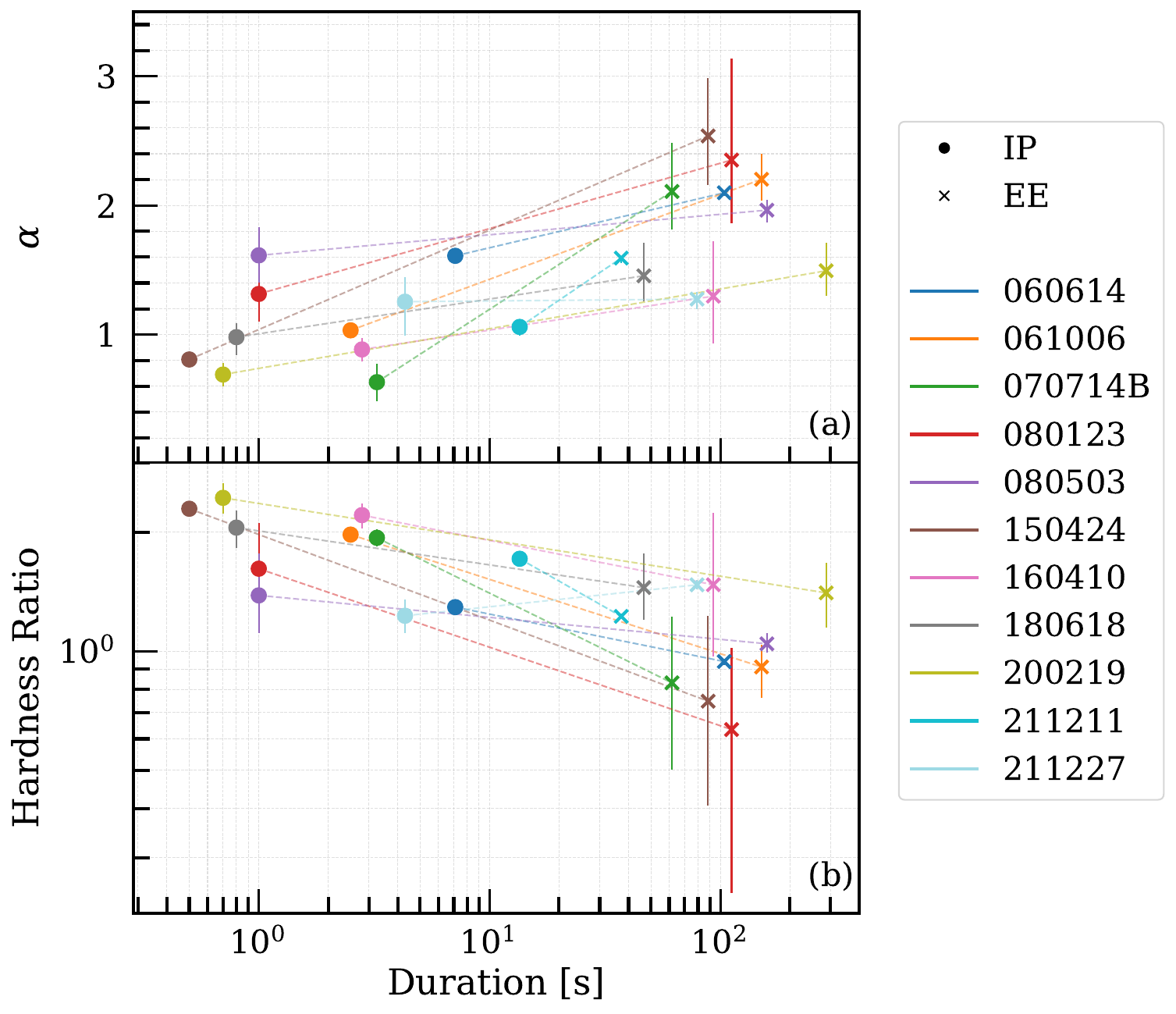}
    \caption{(a): Photon index (panel a) and hardness ratio (panel b) between the IP (dot marker) and EE (cross marker) phases versus the duration the relative phase. Dotted lines show the parameter evolution evolution across a single burst (each in a different color).}
    \label{fig:ph_hr}
\end{figure}

Here we present the results for the prompt emission analysis. In Table~\ref{tab:results} we report the $T_{\text{90}}$ and the duration of the two phases, $T_{IP}$ and $T_{EE}$. The spectral photon indices obtained for both phases are summarized in Table~\ref{tab:results} and in Fig.~\ref{fig:ph_hr}. 
\par Based on the temporal properties of the prompt light curves, SGRBEEs in our sample can be divided into two main categories: \textit{pulse-like} and \textit{tail-like}. In the former, the EE appears as a distinct secondary emission episode following a hard initial pulse. In the latter, the EE is largely background dominated; however, considering the cumulative function, the accumulated signal is enough to extend the duration of the event. 

As shown in Fig.~\ref{fig:ph_hr} (top panel), the spectral evolution of the bursts in our sample generally follows an hard-to-soft trend. For each GRB, $\alpha$ tends to increase when transitioning from the IP to the EE phase, indicating that the spectrum becomes softer. This trend is consistently supported by the hardness ratio analysis (Fig.~\ref{fig:ph_hr}, bottom panel), which shows a general decrease from the IP to the EE. Among the events in our sample, only GRB 211227A does not clearly follow the hard-to-soft evolution observed both in $\alpha$ and hardness ratio trends. In order to better understand the spectral behavior of GRB 211227A, we performed a more detailed analysis of the light curve in different energy bands and a time-resolved analysis of the spectrum, dividing the EE into smaller time intervals. This analysis revealed a more complex spectral evolution than the one we obtain from analysing the whole EE phase. After the hard IP, the spectrum enters a phase in which $\alpha$ becomes lower (i.e. harder) than that of the IP for approximately $30$ s. After, $\alpha$ starts to increase again, indicating a progressive spectral softening, as also reported by \cite{Lu_2022}. This suggests that the spectral evolution of the burst is characterized by an initial hard spike, followed by a relatively hard intermediate phase, and finally by a softer EE phase.

It is worth mentioning that it is not possible to define univocal reference values of $\alpha$ and $HR$ that allow to determine whether a spectrum is typically hard or soft.
In this work, we generally observe a hard-to-soft evolution in single events. However, HR values during the IP phase of some GRBs might be equivalent to the ones during the EE phase of other sources (see Fig.~\ref{fig:ph_hr}).

\begin{table*}
\centering
\caption{Temporal and spectral properties of the SGRBEE sample in the extended SBAT$4$ catalog.}
\label{tab:results}
\begin{tabular}{ccccccccc}
\toprule\toprule

GRB & $z$ & $T_{\text{90}}$ & Phase & $T_{\text{phase}}$ & $\Gamma$ & $F_{15-150}$ & $E_{\text{peak}}$ & $HR$ \\[0.5ex]
 &  & [s] &  & [s] &  & [$10^{-8}$ erg/cm$^2$/s] & [keV] &  \\

\midrule

\multirow{2}{*}[-0.5ex]{060614} & \multirow{2}{*}[-0.5ex]{0.125} & \multirow{2}{*}[-0.5ex]{111.11} & IP & 7.10 & 
$1.61_{-0.04}^{+0.04}$ & $45.92_{-1.23}^{+1.11}$ & -- & $1.29_{-0.04}^{+0.06}$ \\[1ex]
 &  &  & EE & 104.01 & 
$2.10_{-0.03}^{+0.02}$ & $14.61_{-0.15}^{+0.18}$ & -- & $0.94_{-0.02}^{+0.01}$ \\[1ex]

\midrule

\multirow{2}{*}[-0.5ex]{061006} & \multirow{2}{*}[-0.5ex]{0.461} & \multirow{2}{*}[-0.5ex]{153.27} & IP & 2.50 & 
$1.03_{-0.05}^{+0.06}$ & $22.82_{-0.72}^{+0.55}$ & -- & $1.97_{-0.09}^{+0.07}$ \\[1ex]
 &  &  & EE & 150.77 & 
$2.20_{-0.16}^{+0.19}$ & $0.56_{-0.05}^{+0.04}$ & -- & $0.91_{-0.15}^{+0.13}$ \\[1ex]

\midrule

\multirow{2}{*}[-0.5ex]{070714B} & \multirow{2}{*}[-0.5ex]{0.923} & \multirow{2}{*}[-0.5ex]{64.94} & IP & 3.25 & 
$0.63_{-0.15}^{+0.14}$ & $15.11_{-0.49}^{+0.49}$ & $186_{-32}^{+46}$ & $1.94_{-0.10}^{+0.10}$ \\[1ex]
 &  &  & EE & 61.69 & 
$2.11_{-0.29}^{+0.38}$ & $0.28_{-0.07}^{+0.08}$ & -- & $0.83_{-0.33}^{+0.39}$ \\[1ex]

\midrule

\multirow{2}{*}[-0.5ex]{080123} & \multirow{2}{*}[-0.5ex]{0.495} & \multirow{2}{*}[-0.5ex]{112.83} & IP & 1.00 & 
$1.32_{-0.21}^{+0.21}$ & $12.28_{-1.73}^{+2.67}$ & -- & $1.62_{-0.40}^{+0.49}$ \\[1ex]
 &  &  & EE & 111.83 & 
$2.35_{-0.49}^{+0.79}$ & $0.25_{-0.07}^{+0.01}$ & -- & $0.63_{-0.39}^{+0.38}$ \\[1ex]

\midrule

\multirow{2}{*}[-0.5ex]{080503} & \multirow{2}{*}[-0.5ex]{--} & \multirow{2}{*}[-0.5ex]{160.08} & IP & 1.00 & 
$1.61_{-0.21}^{+0.22}$ & $4.82_{-0.72}^{+0.67}$ & -- & $1.38_{-0.27}^{+0.38}$ \\[1ex]
 &  &  & EE & 159.08 & 
$1.96_{-0.09}^{+0.08}$ & $1.09_{-0.05}^{+0.05}$ & -- & $1.04_{-0.05}^{+0.07}$ \\[1ex]

\midrule

\multirow{2}{*}[-0.5ex]{150424A} & \multirow{2}{*}[-0.5ex]{1.00} & \multirow{2}{*}[-0.5ex]{89.00} & IP & 0.50 & 
$0.81_{-0.03}^{+0.03}$ & $214.67_{-5.05}^{+5.59}$ & -- & $2.29_{-0.08}^{+0.08}$ \\[1ex]
 &  &  & EE & 88.50 & 
$2.54_{-0.38}^{+0.45}$ & $0.27_{-0.06}^{+0.06}$ & -- & $0.75_{-0.34}^{+0.48}$ \\[1ex]

\midrule

\multirow{2}{*}[-0.5ex]{160410A} & \multirow{2}{*}[-0.5ex]{1.718} & \multirow{2}{*}[-0.5ex]{96.00} & IP & 2.80 & 
$0.89_{-0.09}^{+0.09}$ & $21.77_{-0.76}^{+1.12}$ & -- & $2.21_{-0.16}^{+0.16}$ \\[1ex]
 &  &  & EE & 93.20 & 
$1.30_{-0.37}^{+0.43}$ & $0.41_{-0.10}^{+0.08}$ & -- & $1.47_{-0.50}^{+0.77}$ \\[1ex]

\midrule

\multirow{2}{*}[-0.5ex]{180618A} & \multirow{2}{*}[-0.5ex]{0.52} & \multirow{2}{*}[-0.5ex]{47.42} & IP & 0.80 & 
$0.98_{-0.14}^{+0.11}$ & $23.82_{-1.47}^{+1.33}$ & -- & $2.05_{-0.23}^{+0.22}$ \\[1ex]
 &  &  & EE & 46.62 & 
$1.46_{-0.20}^{+0.25}$ & $0.97_{-0.12}^{+0.11}$ & -- & $1.45_{-0.25}^{+0.32}$ \\[1ex]

\midrule

\multirow{2}{*}[-0.5ex]{200219A} & \multirow{2}{*}[-0.5ex]{0.48} & \multirow{2}{*}[-0.5ex]{288.00} & IP & 0.70 & 
$0.69_{-0.09}^{+0.09}$ & $30.81_{-1.73}^{+1.24}$ & -- & $2.44_{-0.21}^{+0.22}$ \\[1ex]
 &  &  & EE & 287.30 & 
$1.49_{-0.20}^{+0.22}$ & $0.31_{-0.03}^{+0.04}$ & -- & $1.40_{-0.25}^{+0.27}$ \\[1ex]

\midrule

\multirow{2}{*}[-0.5ex]{211211A} & \multirow{2}{*}[-0.5ex]{0.076} & \multirow{2}{*}[-0.5ex]{50.70} & IP & 13.50 & 
$1.06_{-0.07}^{+0.04}$ & $562.06_{-3.46}^{+3.78}$ & $309_{-62}^{+42}$ & $1.71_{-0.02}^{+0.02}$ \\[1ex]
 &  &  & EE & 37.20 & 
$1.59_{-0.04}^{+0.02}$ & $167.96_{-1.02}^{+1.23}$ & $166_{-18}^{+19}$ & $1.23_{-0.01}^{+0.02}$ \\[1ex]

\midrule

\multirow{2}{*}[-0.5ex]{211227A} & \multirow{2}{*}[-0.5ex]{0.228} & \multirow{2}{*}[-0.5ex]{83.51} & IP & 4.30 & 
$1.25_{-0.26}^{+0.19}$ & $15.83_{-1.12}^{+0.66}$ & $82_{-16}^{+29}$ & $1.23_{-0.12}^{+0.12}$ \\[1ex]
 &  &  & EE & 79.21 & 
$1.27_{-0.07}^{+0.06}$ & $8.45_{-0.15}^{+0.14}$ & $220_{-59}^{+58}$ & $1.47_{-0.04}^{+0.04}$ \\[1ex]

\bottomrule\bottomrule

\end{tabular}
\tablefoot{$T_{IP}$ corresponds to the duration of the IP, while $T_{EE}$ is the duration of the EE computed as the difference between $T_{\text{90}}$ and $T_{IP}$. We report the values of the spectral parameters obtained from the best-fit model: $\Gamma$ is the photon index, $F_{15-150}$ is the flux in the 15-150 keV band, and $HR$ is the hardness ratio. We report the peak energy, $E_{peak}$, only when the cutoff power law model is the best-fit model.}
\end{table*}

\par
\begin{figure}
    \centering
    \includegraphics[width=\linewidth]{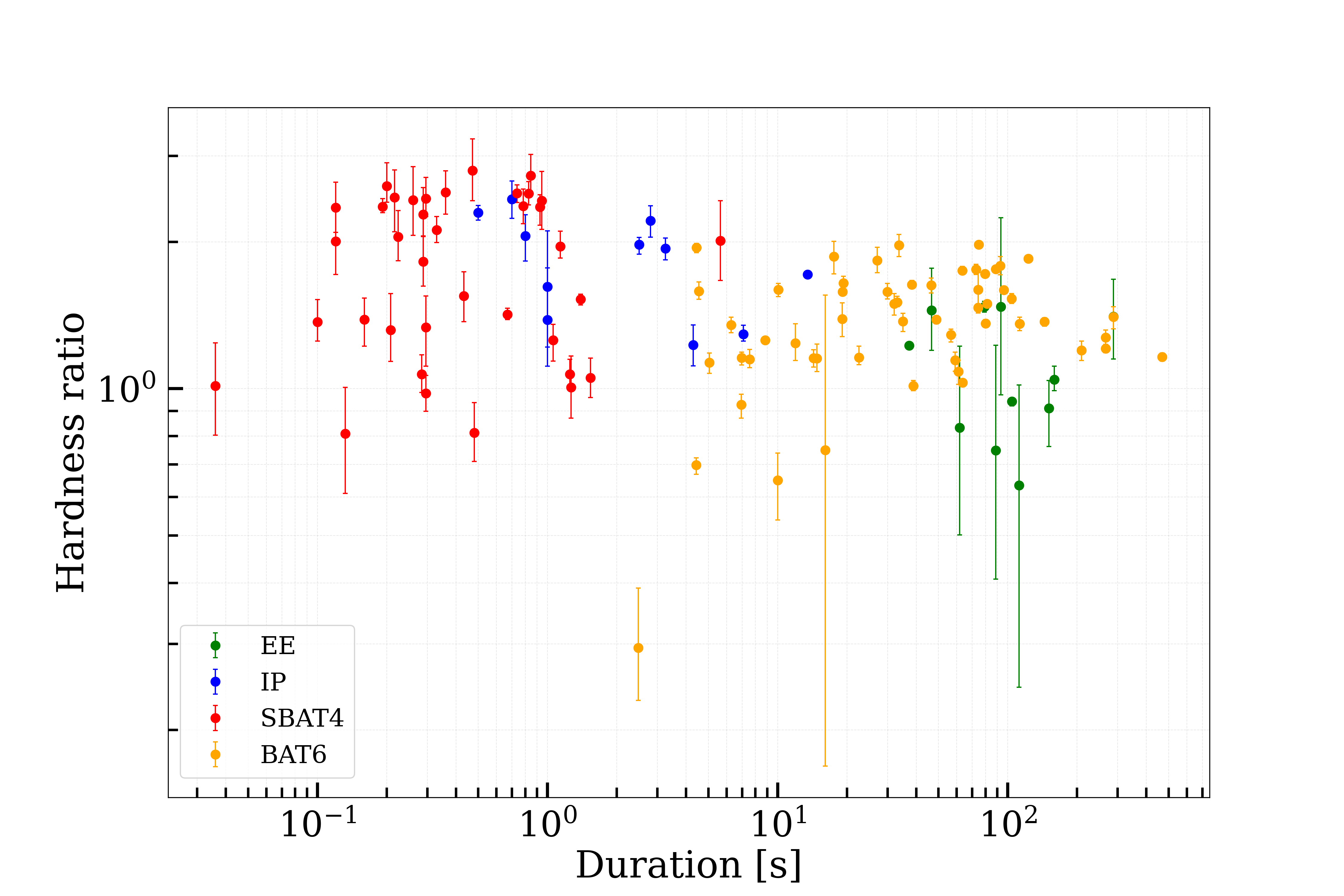}
    \caption{ Hardness ratio ($HR$) versus the burst duration during the IP (in blue) and EE (in green) phases, as well as the total duration of short and long GRBs in the SBAT4E (in red) and BAT6 (in yellow) samples, respectively.}
    \label{fig:hr_all}
\end{figure}

 In order to compare both the IP and EE phases within SGRBs and LGRBs populations, we repeated the same analysis performed for GRBs in the BAT$6$ and extended SBAT4 samples (\citealt{Salvaterra12}, D'Avanzo in preparation) and show the results in Fig.~\ref{fig:hr_all}. 

\par We tested whether the IP and EE durations are correlated with those of SGRBs and LGRBs, respectively, through the Kolmogorov-Smirnov (KS) test. This test assesses whether two apparently independent samples originate from the same probability distribution by measuring the KS value, defined as the maximum distance between their cumulative distribution functions.
We adopted a significance level of $p = 0.05$. We tested the null hypothesis, namely, that the two populations are drawn from the same distribution. If the p-value in a given time is below the significance level, the null hypothesis is rejected and the two populations are independent.
\par The duration distributions already show noticeable differences at a descriptive level: the EE component has a longer average duration ($\overline{T}_{EE}= 110.85 \pm69.92$ s) than the BAT$6$ sample ($\overline{T}_{BAT6}= 65.18 \pm86.23$ s), while the IP is on average longer ($\overline{T}_{IP}= 3.40 \pm3.89$ s) than the extended SBAT4 sample ($\overline{T}_{SBAT4}= 0.67 \pm0.93$ s), although both belong to the short-duration regime.
\par When comparing the EE durations with those of LGRBs in the BAT$6$ sample, we obtain  $KS = 0.49$ and $p = 0.02$. Similarly, when comparing the IP durations with those of SGRBs in the extended SBAT4 sample, we obtain  $KS = 0.62$ and $p = 0.01$. These results indicate that both IP and EE durations are statistically different from those of SGRBs and LGRBs, respectively. 

We repeated the KS test for the hardness ratio distributions, leading to more complex results. We obtained 
$KS = 0.44$ and $p = 0.048$ when comparing $HR$ values from the BAT6 to those during the EE in our sample. The KS value is slightly below the adopted threshold. Therefore, the null hypothesis is formally rejected, although the two distributions can be considered only marginally inconsistent. In contrast, when we compared $HR$ values during IP versus those from the extended SBAT4 sample, we obtain  $KS = 0.29$ and $p = 0.40$, meaning that we can not reject the null hypothesis and the two distribution can be considered as statistically compatible. 

\par In conclusion, SGRBEEs spectra tend to become softer when transitioning from the IP to the EE. This result is consistent with the well-established observation that SGRBs have harder spectra than LGRBs \citep{ghirlanda_2009}. Taken together, the KS results suggest that the IP shares a common physical origin with canonical SGRBs, while the spectral behavior of the EE remains less clear-cut: although only marginally inconsistent with the BAT6 distribution, this hints at a possible connection with the softer emission typically seen in LGRBs.

\subsection{$E_{\text{peak,z}}-E_{\text{iso}}$ correlation}\label{sec:amati}
The $E_{\text{peak, z}}-E_{\text{iso}}$ correlation is a well established empirical scaling law connecting the rest frame spectral peak energy, $E_{\text{peak,z}}$, and the isotropic equivalent radiated energy of the prompt emission, $E_{\text{iso}}$ \citep{Amati_2002, Amati_2008}.
In particular, LGRBs follow a tight power law relation $E_{\text{peak, z}} \propto E_{\text{iso}}^m$, where $m \sim 0.53$ is the power law slope. LGRBs data points dispersion around the best-fit power law is computed through the intrinsic scatter, $\sigma_{sc}$, accounting for systematic variations among different bursts. Therefore, the $3\sigma_{sc}$ scatter region defines the area where approximately 99.7$\%$ of standard LGRBs are expected to lie \citep{Nava_2012}.
SGRBs, on the other hand, populate a different region of the plane: they follow a track that is roughly parallel to the LGRBs one but systematically shifted upwards \citep{Davanzo14}. This implies that, for a given peak energy, a SGRB will have a systematic lower isotropic energy than a LGRB. Therefore, since short and long GRBs populate different regions of the Amati plane, this relation can be used as a diagnostic tool for GRB classification.

\par In this work, we use the Amati plane to investigate the energetic properties during the IP and EE phases separately, comparing our sample with the BAT6 and extended SBAT4 results reported in \cite{Nava_2012} and D'Avanzo in preparation, respectively.

Some events exhibit a spectral peak falling outside the narrow energy range of BAT ($15-150$ keV), leading to unconstrained peak energy values. Therefore,  we performed a joint spectral analysis incorporating data from \textit{Fermi}/GBM ($8$ keV - $40$ MeV) and \textit{Konus}/WIND ($\sim20$ keV–20 MeV), following the fitting routine reported in Section~\ref{sec:fit_routine}. Their broader energy coverage allowed us to observe the full spectral shape and provided a reliable estimate of $E_{\rm peak,z}$ for most of the events with simultaneous observation. 

\begin{table*}
\centering
\caption{Results of the joint spectral analysis for the SGRBEEs with simultaneous observation. }
\label{tab:joint_results}
\begin{tabular}{ccccccccc}
\toprule\toprule

GRB & $z$ & Instrument & Phase & $T_{\rm phase}$ & $E_{\text{peak, z}}$ [keV] & $E_{\text{iso}}$ [$10^{50}$ erg] \\

\midrule

060614  & 0.125 & K      & IP & 8.448 & $375^{+119}_{-114}$ & $4.52^{+0.46}_{-0.56}$ \\[1ex]
\midrule

061006  & 0.461 & K      & IP & 0.256 & $832^{+163}_{-114}$ & $13.92^{+1.85}_{-1.79}$ \\[1ex]
\midrule

150424A & 1.00  & K      & IP & 0.256 & $2243^{+178}_{-134}$ & $497.87^{+33.08}_{-47.42}$ \\[1ex]
\midrule

160410A & 1.718 & K      & IP & 8.448 & $5399^{+1}_{-1344}$ & $1263.10^{+1}_{-143.49}$ \\[1ex]
\midrule

180618A & 0.52  & K - F  & IP & 0.256 & $2919^{+117}_{-192}$ & $27.49^{+1.98}_{-2.33}$ \\[1ex]
\midrule

200219A & 0.48  & K - F  & IP & 0.256 & $1855^{+13}_{-2}$ & $30.03^{+0.10}_{-1.13}$ \\[1ex]
\midrule

\multirow{2}{*}[-0.5ex]{211211A} & \multirow{2}{*}[-0.5ex]{0.228} & \multirow{2}{*}[-0.5ex]{F} 
& IP & 13.5  & $905^{+43}_{-40}$ & $83.93^{+1.04}_{-1.03}$ \\[1ex]

 &  &  & EE & 37.20 & $206^{+23}_{-17}$ & $22.8^{+0.58}_{-0.56}$ \\[1ex]

\midrule

\end{tabular}
\tablefoot{We report results obtained from the Band function fit. We have listed the instruments used for the joint analysis (F: \textit{Fermi}; K: Konus-Wind). $T_{\rm phase}$ is the duration of initial peak (IP) or extended emission (EE) phases. We reported the energies $E_{\text{peak, z}}$ and $E_{\text{iso}}$ obtained from the spectral analysis and shown in Fig.~\ref{fig:amati}.}
\end{table*}

\begin{figure}
\centering
    \centering
\includegraphics[width=\linewidth]{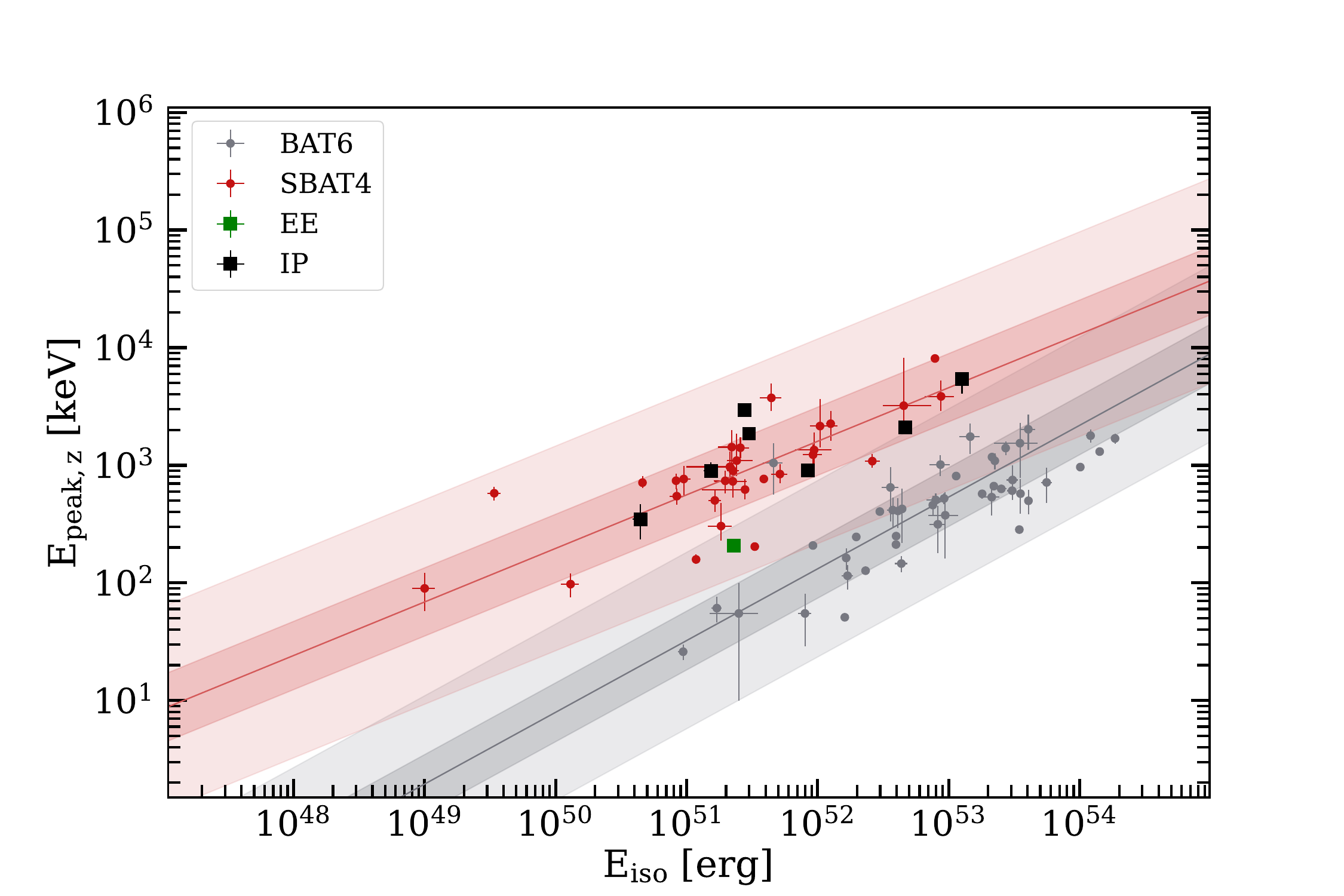}
    \caption{$E_{\text{peak,z}}-E_{\text{iso}}$ (Amati) relation. SGRBs of the extended SBAT4 sample are represented by red dots, shaded regions indicate the $1$ and $3\sigma_{sc}$ scatter of the sample around the best fit (D'Avanzo et al., paper in preparation). LGRBs of the BAT$6$ sample and their fit are represented in grey \citep{Nava_2012}. The power-law best fits are shown as a solid line. Results of the IP and EE joint analyses are shown as black and green squares, respectively.}
    \label{fig:amati}
\end{figure}
\par We report the results in Fig.~\ref{fig:amati}. 
With the exception of GRB 060614 and GRB 211211A, all other events with simultaneous observations exhibit a \textit{tail-like} extended emission. In these GRBs, the spectrum during the EE phase is dominated by the background, leading to unconstrained values of $E_{\text{peak}}$. The inclusion of GBM and Konus-Wind data in the spectral fits enabled the estimation of $E_{\text{peak,z}}$ only during the IP phase, with the only exception of GRB 211211A. \par As shown in Fig.~\ref{fig:amati}, the EE component falls into the overlapping region identified by the $3\sigma$ scatter of the relation found for LGRBs and SGRBs. We must point out that the scatter of the Amati relation is particularly broad ($3\sigma$). Since this correlation is significantly dominated by systematic uncertainties, drawing a physical association based on a component falling within this region should be approached with caution. Nevertheless, it is striking that none of the IPs falls within this broad LGRB region, but rather occupies the SGRB region. The fact that IPs distribute outside the LGRB $3\sigma$ scatter region strongly indicates a different behavior from LGRBs. Overall, this energetic division is consistent with the discussion in Section~\ref{sec:prompt_res}: the IP exhibits spectral properties similar to those of SGRBs, whereas the EE shares characteristics with LGRBs.

\subsection{Spectral lag}\label{sec:lag_res}
\begin{figure}
    \centering
    \includegraphics[width=\linewidth]{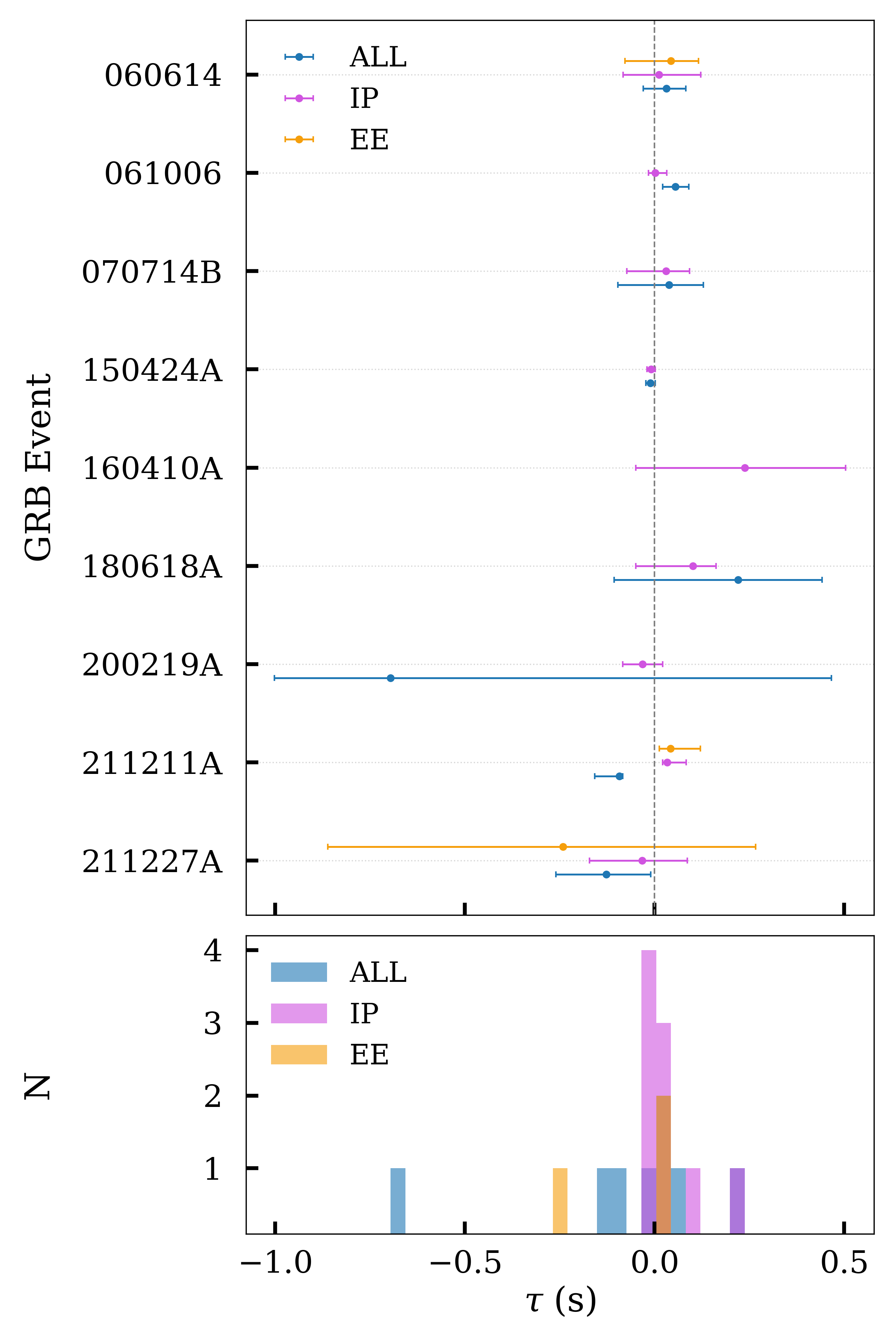}
    \caption{Spectral lag for SGRBEEs in the rest frame energy bands $25-50$ keV and $50-100$ keV considering the whole prompt duration (in blue), the IP (in pink), and the EE (in orange). We report only spectral-lag estimates derived from spectra with enough SNR, as described in Sect.\ref{sec:lag}.}
    \label{fig:lag_sgrbee}
\end{figure}

Due to the intrinsically low SNR characterizing the EE, extracting reliable spectral lags for this phase was not possible for the majority of GRBs in our sample. Thus, to ensure a sufficient SNR for a robust temporal analysis, we adopted different energy bands from those used in \cite{Bernardini_2015}. As previously noted, the EE in \textit{tail-like} GRBs is completely dominated by the background, leading to a low SNR. Consequently, the maximum cross-correlation coefficient for these events failed to meet the \cite{Ukwatta_2010} threshold. A correlation peak below this value implies that the intrinsic source signal is too weak to be distinguished from the background noise. Moreover, for such events, Monte Carlo simulations produced highly irregular or uniformly flat lag distributions. For these reasons, the corresponding results have been omitted from the Fig.~\ref{fig:lag_sgrbee}. Despite the methodological differences in energy band selection, our analysis shows that, with the exception of GRB 211211A, the rest-frame spectral lags for all the analysed events during the IP, the EE and the whole prompt phase are consistent with zero within a $2\sigma$ confidence level (Fig.~\ref{fig:lag_sgrbee}). From this analysis, we find that the spectral lag cannot be used as a key parameter to distinguish between short and long GRBs, as already pointed out by \cite{Bernardini_2015}, nor can it provide a reliable criterion for determining the classification of GRBs exhibiting extended emission.

\subsection{XRT temporal analysis} \label{sec:xrt_res}
\begin{figure*}
    \centering
    \centering
    \includegraphics[width=\textwidth]{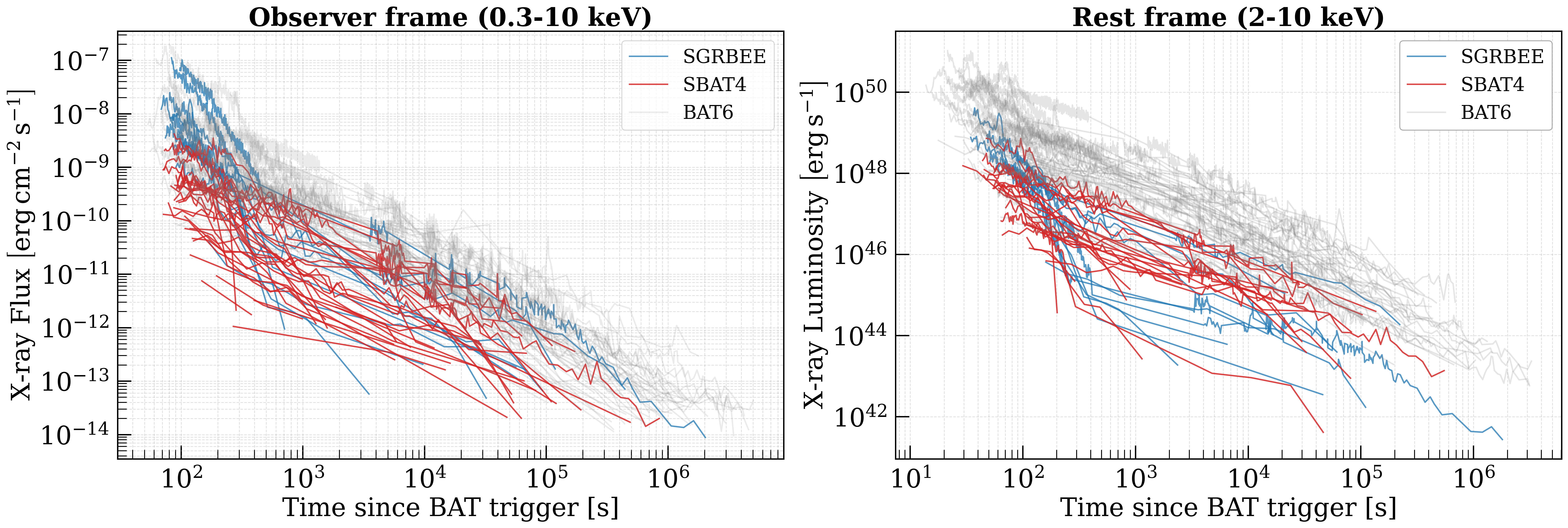}
    \caption{XRT light curves in observer frame (left panel) and in the common rest frame (right panel) of GRBs from the SBAT$4$ sample (in red), BAT6 sample (in grey) and for our sample of GRBs with extended emission (in blue).}
    \label{fig:combined_plot}
\end{figure*}

We studied the XRT light curves both in the observer frame and in the $2-10$ keV common rest frame (Eq.~\ref{eq:common_frame}) for GRBs in both the extended SBAT4 and BAT6 samples \citep{Davanzo_12}. In the observer frame (Fig.~\ref{fig:combined_plot}, left panel) the BAT6 sample of LGRBs exhibits a higher X-ray flux than standard extended SBAT4 SGRBs across the entire temporal window. On the other hand, SGRBEEs show an intermediate behavior. At early times, they exhibit higher fluxes compared to standard extended SBAT4 GRBs, visibly overlapping with those of LGRBs. At later times, beyond $\sim 400$ s, such distinction between SGRBEEs and standard SGRB vanishes. 

Conversely, in the common rest frame (Fig.~\ref{fig:combined_plot}, right panel), differences between X-ray light curves of SGRBs with and without the EE appear less evident. To exclude artifacts due to different redshift distributions, we performed a KS test. The test confirmed that the redshifts of the extended SBAT4 ($z = 0.87 \pm 0.58$) and SGRBEE ($z = 0.66 \pm 0.49$) samples are drawn from compatible distributions. Additionally, our analysis confirms the well-established result that LGRBs are systematically more luminous than SGRBs (\citealt{margutti2012}, Brivio et al., in preparation), but also reveals that SGRBEEs exhibit a similar trend (Fig.~\ref{fig:combined_plot}, right panel). However, we note that there is a slight overlap between the brightest SGRBs and the faintest LGRBs in the common rest frame. 

\begin{figure}
    \centering
    \includegraphics[width=\linewidth]{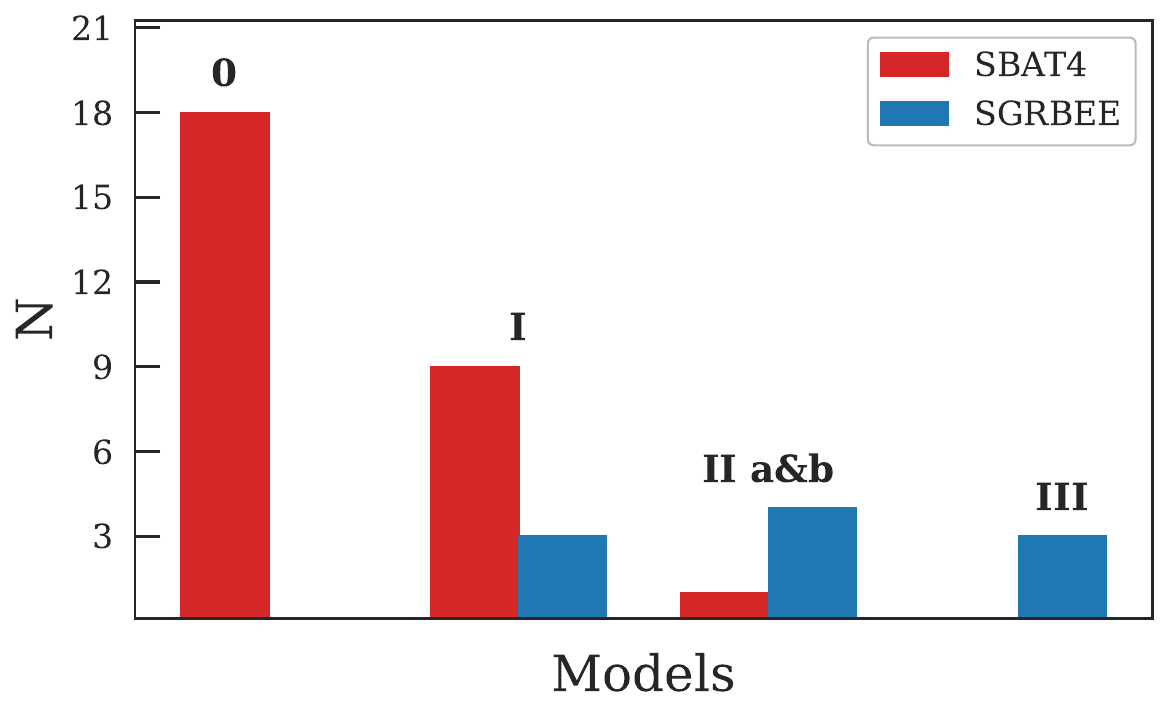}
    \caption{Best-fit model histogram for SBAT4 (in red) and SGRBEEs (in blue) XRT light curves.}
    \label{fig:hist_model}
\end{figure}

To analyse X-ray light curves in more detail, we characterized their morphology following the methodology described in Section~\ref{sec:xrt}. We report the best-fit model histogram for both extended SBAT4 and SGRBEEs populations (Fig.~\ref{fig:hist_model}). First, we observe that only one SGRB of the SBAT4 sample requires a complex model (type II), GRB $231117$A, and none of the SGRBEEs can be described by a single power law model (type 0). This suggests that SGRBEEs exhibit more complex light curves, characterized by multiple temporal breaks. Furthermore, we point out that SGRBs light curves have a lower temporal coverage than SGRBEEs, which naturally tends to favor simpler phenomenological descriptions. However, SGRBEEs often exhibit observational gaps between early time WT mode and late time PC mode data. This introduces a significant degree of degeneracy in the light curve reconstruction. In these intervals, the lack of data means that multiple functional forms can in principle describe the evolution of the emission equally well, since the true behavior of the source is unconstrained. As a consequence, while the late-time points can drive the fit towards more complex models by connecting widely separated temporal segments, such complexity should be interpreted with caution: it is a consequence of sparse sampling. The resulting model should therefore be regarded as reliable only within the time intervals actually covered by observations, while remaining intrinsically non-unique in the gaps between them. We include all available data in our morphological analysis, but we explicitly acknowledge that the reconstruction across observational gaps is not uniquely constrained by the data.

\begin{figure*}
    \centering
    \centering
    \includegraphics[width=\textwidth]{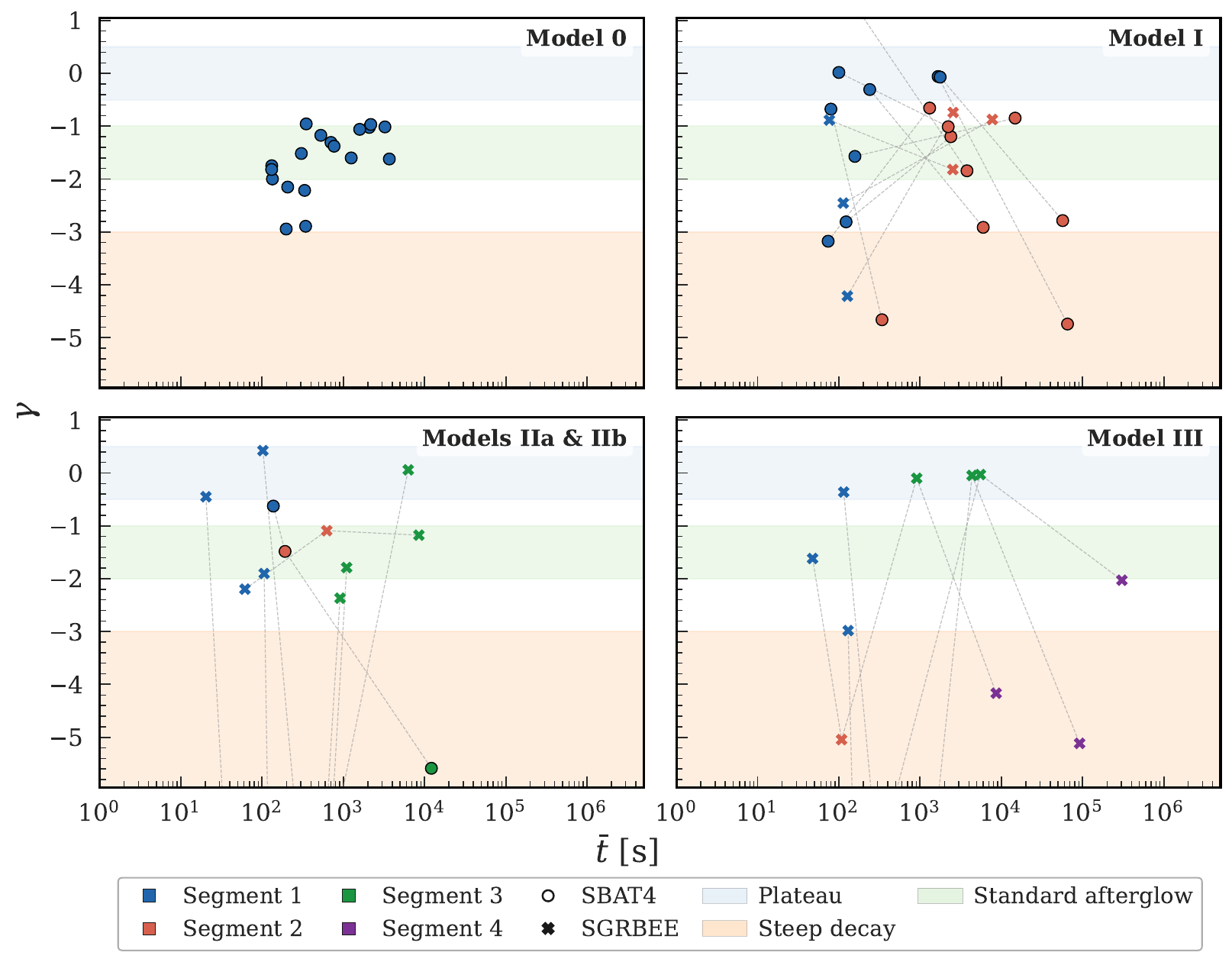}
    \caption{Light curve temporal indices $\gamma$ from the morphological characterization of SBAT$4$ GRBs (circles) and the SGRBEEs (crosses) for each light curve model. $\bar t$ is the average time of the light curve segment characterized by the given temporal index $\gamma$. Each of the light curve segments are shown with different colors. As a reference, we include the expected $\gamma$ values for the plateau, steep decay, and standard afterglow phases in blue, red and green, respectively \citep{Zhang_2007}.}
    \label{fig:morph_res}
\end{figure*}

In Fig.~\ref{fig:morph_res} we reported the $\gamma$ parameter found for each model. We can summarize the results as follows:
\begin{itemize}
    \item \textbf{Model 0:} Only SGRBs with a sparsely sampled X-ray light curve are best fitted by Model 0, i.e. a single power law segment. For these specific events, the lack of sufficient data dictates the use of Model 0, as it prevents overfitting and adequately describes the limited available observations. Several SGRBs exhibit temporal indices consistent with values in between those predicted by the afterglow model before and after a possible jet break ($-2 < \gamma < -1$; \citealt{Sari1998, Sari1999, Rhoads1999, Frail2001}). However, this behavior is not observed for all the GRBs, with some of them exhibiting faster decays with respect to standard afterglow predictions.\\

    \item \textbf{Model I:} Light curves described by this model show the presence of a temporal break, indicating a variation in the temporal index during the afterglow emission. Interestingly, 2/3 SGRBEEs in this category feature an initial steep decay phase ($\gamma < -3$). We can classify these into four distinct morphological sub-categories:
    \begin{itemize}
        \item[\textbf{A}.)] Light curves characterized by an initial plateau phase ($ -0.5<\gamma< 0.5$) followed by a subsequent decay;
        \item[\textbf{B}.)] Light curves in this sub-category tend to become more shallow during the transition from the first to the second segment. Notably, one GRB in this sub-category, GRB $131004$A, exhibits a positive temporal index during its initial phase (i.e. a rising flux, $\gamma > 0$) driven by an early flux excess. The automated fitting routine did not classify this initial rising feature as a flare. Most probably this initial flux rising is due to the afterglow coasting phase;
        \item[\textbf{C}.)] Light curves exhibiting a canonical afterglow decay ($\gamma \sim -1$), followed by a steepening with large temporal indices ($\gamma < -1.5$);
        \item[\textbf{D}.)] Light curves which begin with an initial steep decay phase that subsequently flattens into a canonical afterglow decay.
    \end{itemize}
    
    \item \textbf{Model II:} 
    Light curves described by this model show the presence of two temporal breaks, highlighting a complex morphology which cannot be explained by mean of the standard afterglow theory. Only one SGRB standard X-ray light curve is best fitted by this model. In particular, two different functions belong to this class (see Eqs.~\ref{eq:XLC_modelIIa} and \ref{eq:XLC_modelIIb}). Three GRBs in this group follow a shallow-steep-shallow decay morphology (model IIb). One event with EE, GRB $160410$A, shows an initial rise followed by a decay (model IIa). One standard SGRB, GRB $231117$A, has model IIa as best fit model.
    \item \textbf{Model III:} similarly to other highly structured light curves in our sample, all events best fitted by this model belong to the SGRBEE class. Morphologically, two out of three events begin with a decay steeper than the canonical afterglow, which then evolves into a steep decay phase before flattening out. Crucially, all three events in this category ultimately exhibit a distinct plateau phase.
\end{itemize}

We further report that an investigation of the available optical data, presented in Brivio et al., in preparation, yielded no significant statistical distinction between SGRBEEs and standard SGRBs.
\par These results suggest that extended emission predominantly acts as an X-ray energy injection mechanism, shaping the temporal evolution of the light curve. SGRBEEs exhibit significantly higher morphological complexity in their X-ray emission than standard SGRBs, likely reflecting prolonged central engine activity, while no significant differences are observed in the optical band. We also note that SGRBEEs appear more luminous than SGRBs at very early times (Fig.~\ref{fig:combined_plot}, right panel), which may indicate an additional contribution from sustained central engine activity or more efficient early-time energy injection. Despite this difference in temporal structure, the absence of a clear luminosity separation in the common rest frame suggests that the overall X-ray energetics of the two populations are comparable.

\section{Conclusions}\label{sec:concl}
In this paper, we presented a comprehensive multi-wavelength characterization of the gamma-ray prompt emission and X-ray afterglow of GRBs with EE observed by \textit{Swift}. These type of events have been widely debated, since the classification based solely on the duration of the prompt emission has been strongly challenged by recent events such as GRB 211211A and GRB 230307A \citep{Rastinejad_22, Troja_22, Levan_24}. Starting from the flux-limited and redshift complete extended SBAT4  and BAT6 samples, we compared SGRBEEs with the standard short and long GRB population. We aim to determine whether SGRBEEs represent a distinct class, an intermediate population, or an extreme manifestation of standard compact binary mergers. 
\par To address the question of whether SGRBEEs represent a distinct population or belong to the standard SGRB class, our multi-wavelength findings reveal a complex scenario where different observables yield contrasting indications. Starting from a temporal perspective, we identified two different categories of EE: \textit{pulse-like} and \textit{tail-like}. While the former temporally resemble standard prompt emission pulses, the latter are hardly distinguishable from the background, making it difficult to properly characterize their spectra. Crucially, KS tests on the prompt emission demonstrate that both IP and EE durations are statistically different, and longer, than those of standard short and long GRBs, respectively. It is worth mentioning that, at odds with the historical classification of SGRBs with extended emission, most of the IPs in our sample last more than 2 s. Notably, events with longer IPs tend to cluster in the region of the $HR$-duration diagram occupied by LGRBs (Fig.~\ref{fig:hr_all}). These specific events (GRB 060614, GRB 211211A, and GRB 211227A) exhibit a \textit{pulse-like} morphology.
Despite these temporal differences, the spectral analysis reveals deeper connections to standard populations. We observe a typical hard-to-soft spectral evolution during the transition from IP to EE, consistent with previous studies \citep{Kaneko_2015}, with the sole exception of GRB 211227A, discussed in Section~\ref{sec:prompt_res}. However, IPs and EEs share similar hardness ratios across different bursts, preventing a global definition of IPs as "hard" and EEs as "soft" beyond their intra-burst evolution. KS tests on spectral hardness further confirm this complex picture: IPs are compatible with the SBAT4 sample ($KS = 0.29$ and $p = 0.40$), whereas EEs show poor compatibility with the BAT6 sample ($KS=0.44$, $p=0.048$). 
\par The $E_{\rm peak, z}$-$E_{\rm iso}$ plane provides an additional diagnostic. Previous studies have shown that classical SGRBs typically fall above the Amati relation \citep{Amati_2008, Nava_2012}. In our analysis, the IPs do fall within the scatter region occupied by SGRBs, despite the duration of several of them being more than 2 s (Fig.~\ref{fig:amati}). For the EE, we could constrain $E_{\rm peak, z}$  only for GRB 211211A, which indeed lies in the overlap region between the long and short scatter distributions. The primary limitation here is that most events with joint observations exhibit a \textit{tail-like} morphology. This results in spectra with low SNR, making the estimation of $E_{\rm peak, z}$ particularly challenging. Overall, this suggests that the two phases trace distinct emission mechanisms rather than different progenitors.
Adding to this picture, SGRBEE spectral lags are consistent with zero. As already noted by \cite{Bernardini_2015}, while typical of SGRBs, this behavior is not uniquely discriminatory as similar values are found in LGRB samples. The X-ray afterglow analysis, however, provides a critical diagnostic. In the common rest frame, the total energetics and early light curves of bursts with and without EE are indistinguishable. This standard rest-frame energy budget points to the same progenitor class. Conversely, SGRBEE X-ray light curves are systematically more complex: $90\%$ exhibit multiple temporal breaks (type II and III morphologies), whereas only one standard SGRB in our sample has two breaks. This combination of standard rest-frame energetics and highly structured morphology departing from the standard forward shock model requires a sustained energy injection. Consequently, the rest-frame energetics, spectral lags, and IP spectral properties all link these events to standard compact binary mergers. The EE itself and the complex X-ray afterglow, instead, serve as direct observational evidence of an active, long-lived post-merger central engine.
The main limitation of this study is the relatively small number of events with available broadband prompt data, which restricts the determination of $E_{\rm peak}$ and energetics for both phases, compounded by the intrinsic faintness of the \textit{tail-like} EEs. Despite the fact that the sample was not statistically large, the selection criteria of the SBAT4 sample allow us to extract robust information free from significant observational biases. Future observations with wide-band instruments will be essential to better constrain the spectral properties of the EE and to clarify its physical origin. In this respect, the SVOM mission \citep{Cordier_15, Wei_16}, thanks to its broad energy coverage that extends into the soft X-ray domain (down to $\sim4$ keV with ECLAIRs, \citealt{Godet_26}), will play a crucial role in detecting and characterizing the often soft spectral signatures of the EE, providing tighter constraints on $E_{\rm peak}$  even for events whose emission peaks below the energy range typically accessible to past and current gamma-ray monitors.

\begin{acknowledgements}
This work was supported by funding from the Italian Space Agency, contract ASI/INAF n.I/004/11/6. PDA, RB, MF, RS, SCa and SCo acknowledge support from the INAF grant no. 1.05.24.04.06. MGB, MD, PDA, and SCa acknowledge support from the INAF grant no. 1.05.24.02.23. The work of S.B., D.F., and A.T. was supported by the basic
funding program of the Ioffe Institute FFUG-2024-0002.

\end{acknowledgements}
    
\bibliographystyle{aa}
\bibliography{biblio}

@ARTICLE{Abbott17,
       author = {{Abbott}, B.~P. and {Abbott}, R. and {Abbott}, T.~D. and {Acernese}, F. and {Ackley}, K. and {Adams}, C. and {Adams}, T. and {Addesso}, P. and {Adhikari}, R.~X. and {Adya}, V.~B. and {Affeldt}, C. and {Afrough}, M. and {Agarwal}, B. and {Agathos}, M. and {Agatsuma}, K. and {Aggarwal}, N. and {Aguiar}, O.~D. and {Aiello}, L. and {Ain}, A. and {Ajith}, P. and {Allen}, B. and {Allen}, G. and {Allocca}, A. and {Aloy}, M.~A. and {Altin}, P.~A. and {Amato}, A. and {Ananyeva}, A. and {Anderson}, S.~B. and {Anderson}, W.~G. and {Angelova}, S.~V. and {Antier}, S. and {Appert}, S. and {Arai}, K. and {Araya}, M.~C. and {Areeda}, J.~S. and {Arnaud}, N. and {Arun}, K.~G. and {Ascenzi}, S. and {Ashton}, G. and {Ast}, M. and {Aston}, S.~M. and {Astone}, P. and {Atallah}, D.~V. and {Aufmuth}, P. and {Aulbert}, C. and {AultONeal}, K. and {Austin}, C. and {Avila-Alvarez}, A. and {Babak}, S. and {Bacon}, P. and {Bader}, M.~K.~M. and {Bae}, S. and {Baker}, P.~T. and {Baldaccini}, F. and {Ballardin}, G. and {Ballmer}, S.~W. and {Banagiri}, S. and {Barayoga}, J.~C. and {Barclay}, S.~E. and {Barish}, B.~C. and {Barker}, D. and {Barkett}, K. and {Barone}, F. and {Barr}, B. and {Barsotti}, L. and {Barsuglia}, M. and {Barta}, D. and {Bartlett}, J. and {Bartos}, I. and {Bassiri}, R. and {Basti}, A. and {Batch}, J.~C. and {Bawaj}, M. and {Bayley}, J.~C. and {Bazzan}, M. and {B{\'e}csy}, B. and {Beer}, C. and {Bejger}, M. and {Belahcene}, I. and {Bell}, A.~S. and {Berger}, B.~K. and {Bergmann}, G. and {Bero}, J.~J. and {Berry}, C.~P.~L. and {Bersanetti}, D. and {Bertolini}, A. and {Betzwieser}, J. and {Bhagwat}, S. and {Bhandare}, R. and {Bilenko}, I.~A. and {Billingsley}, G. and {Billman}, C.~R. and {Birch}, J. and {Birney}, R. and {Birnholtz}, O. and {Biscans}, S. and {Biscoveanu}, S. and {Bisht}, A. and {Bitossi}, M. and {Biwer}, C. and {Bizouard}, M.~A. and {Blackburn}, J.~K. and {Blackman}, J. and {Blair}, C.~D. and {Blair}, D.~G. and {Blair}, R.~M. and {Bloemen}, S. and {Bock}, O. and {Bode}, N. and {Boer}, M. and {Bogaert}, G. and {Bohe}, A. and {Bondu}, F. and {Bonilla}, E. and {Bonnand}, R. and {Boom}, B.~A. and {Bork}, R. and {Boschi}, V. and {Bose}, S. and {Bossie}, K. and {Bouffanais}, Y. and {Bozzi}, A. and {Bradaschia}, C. and {Brady}, P.~R. and {Branchesi}, M. and {Brau}, J.~E. and {Briant}, T. and {Brillet}, A. and {Brinkmann}, M. and {Brisson}, V. and {Brockill}, P. and {Broida}, J.~E. and {Brooks}, A.~F. and {Brown}, D.~A. and {Brown}, D.~D. and {Brunett}, S. and {Buchanan}, C.~C. and {Buikema}, A. and {Bulik}, T. and {Bulten}, H.~J. and {Buonanno}, A. and {Buskulic}, D. and {Buy}, C. and {Byer}, R.~L. and {Cabero}, M. and {Cadonati}, L. and {Cagnoli}, G. and {Cahillane}, C. and {Calder{\'o}n Bustillo}, J. and {Callister}, T.~A. and {Calloni}, E. and {Camp}, J.~B. and {Canepa}, M. and {Canizares}, P. and {Cannon}, K.~C. and {Cao}, H. and {Cao}, J. and {Capano}, C.~D. and {Capocasa}, E. and {Carbognani}, F. and {Caride}, S. and {Carney}, M.~F. and {Casanueva Diaz}, J. and {Casentini}, C. and {Caudill}, S. and {Cavagli{\`a}}, M. and {Cavalier}, F. and {Cavalieri}, R. and {Cella}, G. and {Cepeda}, C.~B. and {Cerd{\'a}-Dur{\'a}n}, P. and {Cerretani}, G. and {Cesarini}, E. and {Chamberlin}, S.~J. and {Chan}, M. and {Chao}, S. and {Charlton}, P. and {Chase}, E. and {Chassande-Mottin}, E. and {Chatterjee}, D. and {Chatziioannou}, K. and {Cheeseboro}, B.~D. and {Chen}, H.~Y. and {Chen}, X. and {Chen}, Y. and {Cheng}, H.-P. and {Chia}, H. and {Chincarini}, A. and {Chiummo}, A. and {Chmiel}, T. and {Cho}, H.~S. and {Cho}, M. and {Chow}, J.~H. and {Christensen}, N. and {Chu}, Q. and {Chua}, A.~J.~K. and {Chua}, S. and {Chung}, A.~K.~W. and {Chung}, S. and {Ciani}, G.},
        title = "{Gravitational Waves and Gamma-Rays from a Binary Neutron Star Merger: GW170817 and GRB 170817A}",
      journal = {\apjl},
         year = 2017,
        month = oct,
       volume = {848},
       number = {2},
          eid = {L13},
        pages = {L13},
          doi = {10.3847/2041-8213/aa920c},
archivePrefix = {arXiv},
       eprint = {1710.05834},
 primaryClass = {astro-ph.HE},
       adsurl = {https://ui.adsabs.harvard.edu/abs/2017ApJ...848L..13A}
}

@ARTICLE{Abbott17_2,
       author = {{Abbott}, B.~P. and {Abbott}, R. and {Abbott}, T.~D. and {Acernese}, F. and {Ackley}, K. and {Adams}, C. and {Adams}, T. and {Addesso}, P. and {Adhikari}, R.~X. and {Adya}, V.~B. and {Affeldt}, C. and {Afrough}, M. and {Agarwal}, B. and {Agathos}, M. and {Agatsuma}, K. and {Aggarwal}, N. and {Aguiar}, O.~D. and {Aiello}, L. and {Ain}, A. and {Ajith}, P. and {Allen}, B. and {Allen}, G. and {Allocca}, A. and {Altin}, P.~A. and {Amato}, A. and {Ananyeva}, A. and {Anderson}, S.~B. and {Anderson}, W.~G. and {Angelova}, S.~V. and {Antier}, S. and {Appert}, S. and {Arai}, K. and {Araya}, M.~C. and {Areeda}, J.~S. and {Arnaud}, N. and {Arun}, K.~G. and {Ascenzi}, S. and {Ashton}, G. and {Ast}, M. and {Aston}, S.~M. and {Astone}, P. and {Atallah}, D.~V. and {Aufmuth}, P. and {Aulbert}, C. and {AultONeal}, K. and {Austin}, C. and {Avila-Alvarez}, A. and {Babak}, S. and {Bacon}, P. and {Bader}, M.~K.~M. and {Bae}, S. and {Bailes}, M. and {Baker}, P.~T. and {Baldaccini}, F. and {Ballardin}, G. and {Ballmer}, S.~W. and {Banagiri}, S. and {Barayoga}, J.~C. and {Barclay}, S.~E. and {Barish}, B.~C. and {Barker}, D. and {Barkett}, K. and {Barone}, F. and {Barr}, B. and {Barsotti}, L. and {Barsuglia}, M. and {Barta}, D. and {Barthelmy}, S.~D. and {Bartlett}, J. and {Bartos}, I. and {Bassiri}, R. and {Basti}, A. and {Batch}, J.~C. and {Bawaj}, M. and {Bayley}, J.~C. and {Bazzan}, M. and {B{\'e}csy}, B. and {Beer}, C. and {Bejger}, M. and {Belahcene}, I. and {Bell}, A.~S. and {Berger}, B.~K. and {Bergmann}, G. and {Bernuzzi}, S. and {Bero}, J.~J. and {Berry}, C.~P.~L. and {Bersanetti}, D. and {Bertolini}, A. and {Betzwieser}, J. and {Bhagwat}, S. and {Bhandare}, R. and {Bilenko}, I.~A. and {Billingsley}, G. and {Billman}, C.~R. and {Birch}, J. and {Birney}, R. and {Birnholtz}, O. and {Biscans}, S. and {Biscoveanu}, S. and {Bisht}, A. and {Bitossi}, M. and {Biwer}, C. and {Bizouard}, M.~A. and {Blackburn}, J.~K. and {Blackman}, J. and {Blair}, C.~D. and {Blair}, D.~G. and {Blair}, R.~M. and {Bloemen}, S. and {Bock}, O. and {Bode}, N. and {Boer}, M. and {Bogaert}, G. and {Bohe}, A. and {Bondu}, F. and {Bonilla}, E. and {Bonnand}, R. and {Boom}, B.~A. and {Bork}, R. and {Boschi}, V. and {Bose}, S. and {Bossie}, K. and {Bouffanais}, Y. and {Bozzi}, A. and {Bradaschia}, C. and {Brady}, P.~R. and {Branchesi}, M. and {Brau}, J.~E. and {Briant}, T. and {Brillet}, A. and {Brinkmann}, M. and {Brisson}, V. and {Brockill}, P. and {Broida}, J.~E. and {Brooks}, A.~F. and {Brown}, D.~A. and {Brown}, D.~D. and {Brunett}, S. and {Buchanan}, C.~C. and {Buikema}, A. and {Bulik}, T. and {Bulten}, H.~J. and {Buonanno}, A. and {Buskulic}, D. and {Buy}, C. and {Byer}, R.~L. and {Cabero}, M. and {Cadonati}, L. and {Cagnoli}, G. and {Cahillane}, C. and {Calder{\'o}n Bustillo}, J. and {Callister}, T.~A. and {Calloni}, E. and {Camp}, J.~B. and {Canepa}, M. and {Canizares}, P. and {Cannon}, K.~C. and {Cao}, H. and {Cao}, J. and {Capano}, C.~D. and {Capocasa}, E. and {Carbognani}, F. and {Caride}, S. and {Carney}, M.~F. and {Carullo}, G. and {Casanueva Diaz}, J. and {Casentini}, C. and {Caudill}, S. and {Cavagli{\`a}}, M. and {Cavalier}, F. and {Cavalieri}, R. and {Cella}, G. and {Cepeda}, C.~B. and {Cerd{\'a}-Dur{\'a}n}, P. and {Cerretani}, G. and {Cesarini}, E. and {Chamberlin}, S.~J. and {Chan}, M. and {Chao}, S. and {Charlton}, P. and {Chase}, E. and {Chassande-Mottin}, E. and {Chatterjee}, D. and {Chatziioannou}, K. and {Cheeseboro}, B.~D. and {Chen}, H.~Y. and {Chen}, X. and {Chen}, Y. and {Cheng}, H.-P. and {Chia}, H. and {Chincarini}, A. and {Chiummo}, A. and {Chmiel}, T. and {Cho}, H.~S. and {Cho}, M. and {Chow}, J.~H. and {Christensen}, N. and {Chu}, Q. and {Chua}, A.~J.~K. and {Chua}, S.},
        title = "{GW170817: Observation of Gravitational Waves from a Binary Neutron Star Inspiral}",
      journal = {\prl},
         year = 2017,
        month = oct,
       volume = {119},
       number = {16},
          eid = {161101},
        pages = {161101},
          doi = {10.1103/PhysRevLett.119.161101},
archivePrefix = {arXiv},
       eprint = {1710.05832},
 primaryClass = {gr-qc},
       adsurl = {https://ui.adsabs.harvard.edu/abs/2017PhRvL.119p1101A}
}

@ARTICLE{akaike_74,
       author = {{Akaike}, H.},
        title = "{A New Look at the Statistical Model Identification}",
      journal = {IEEE Transactions on Automatic Control},
         year = 1974,
        month = jan,
       volume = {19},
        pages = {716-723},
          doi = {10.1109/TAC.1974.1100705},
       adsurl = {https://ui.adsabs.harvard.edu/abs/1974ITAC...19..716A}
}

@ARTICLE{amati06,
       author = {{Amati}, Lorenzo},
        title = "{The E$_{p,i}$-E$_{iso}$ correlation in gamma-ray bursts: updated observational status, re-analysis and main implications}",
      journal = {\mnras},
         year = 2006,
        month = oct,
       volume = {372},
       number = {1},
        pages = {233-245},
          doi = {10.1111/j.1365-2966.2006.10840.x},
archivePrefix = {arXiv},
       eprint = {astro-ph/0601553},
 primaryClass = {astro-ph},
       adsurl = {https://ui.adsabs.harvard.edu/abs/2006MNRAS.372..233A}
}

@ARTICLE{Amati_2008,
       author = {{Amati}, Lorenzo and {Guidorzi}, Cristiano and {Frontera}, Filippo and {Della Valle}, Massimo and {Finelli}, Fabio and {Landi}, Raffaella and {Montanari}, Enrico},
        title = "{Measuring the cosmological parameters with the E$_{p,i}$-E$_{iso}$ correlation of gamma-ray bursts}",
      journal = {\mnras},
         year = 2008,
        month = dec,
       volume = {391},
       number = {2},
        pages = {577-584},
          doi = {10.1111/j.1365-2966.2008.13943.x},
archivePrefix = {arXiv},
       eprint = {0805.0377},
 primaryClass = {astro-ph},
       adsurl = {https://ui.adsabs.harvard.edu/abs/2008MNRAS.391..577A}
}

@ARTICLE{Amati_2002,
       author = {{Amati}, L. and {Frontera}, F. and {Tavani}, M. and {in't Zand}, J.~J.~M. and {Antonelli}, A. and {Costa}, E. and {Feroci}, M. and {Guidorzi}, C. and {Heise}, J. and {Masetti}, N. and {Montanari}, E. and {Nicastro}, L. and {Palazzi}, E. and {Pian}, E. and {Piro}, L. and {Soffitta}, P.},
        title = "{Intrinsic spectra and energetics of BeppoSAX Gamma-Ray Bursts with known redshifts}",
      journal = {\aap},
         year = 2002,
        month = jul,
       volume = {390},
        pages = {81-89},
          doi = {10.1051/0004-6361:20020722},
archivePrefix = {arXiv},
       eprint = {astro-ph/0205230},
 primaryClass = {astro-ph},
       adsurl = {https://ui.adsabs.harvard.edu/abs/2002A&A...390...81A}
}

@ARTICLE{Aptekar1995,
       author = {{Aptekar}, R.~L. and {Frederiks}, D.~D. and {Golenetskii}, S.~V. and {Ilynskii}, V.~N. and {Mazets}, E.~P. and {Panov}, V.~N. and {Sokolova}, Z.~J. and {Terekhov}, M.~M. and {Sheshin}, L.~O. and {Cline}, T.~L. and {Stilwell}, D.~E.},
        title = "{Konus-W Gamma-Ray Burst Experiment for the GGS Wind Spacecraft}",
      journal = {\ssr},
         year = 1995,
        month = feb,
       volume = {71},
       number = {1-4},
        pages = {265-272},
          doi = {10.1007/BF00751332},
       adsurl = {https://ui.adsabs.harvard.edu/abs/1995SSRv...71..265A}
}

@ARTICLE{band93,
       author = {{Band}, D. and {Matteson}, J. and {Ford}, L. and {Schaefer}, B. and {Palmer}, D. and {Teegarden}, B. and {Cline}, T. and {Briggs}, M. and {Paciesas}, W. and {Pendleton}, G. and {Fishman}, G. and {Kouveliotou}, C. and {Meegan}, C. and {Wilson}, R. and {Lestrade}, P.},
        title = "{BATSE Observations of Gamma-Ray Burst Spectra. I. Spectral Diversity}",
      journal = {\apj},
         year = 1993,
        month = aug,
       volume = {413},
        pages = {281},
          doi = {10.1086/172995},
       adsurl = {https://ui.adsabs.harvard.edu/abs/1993ApJ...413..281B}
}

@ARTICLE{Evans_10,
       author = {{Evans}, P.~A. and {Willingale}, R. and {Osborne}, J.~P. and {O'Brien}, P.~T. and {Page}, K.~L. and {Markwardt}, C.~B. and {Barthelmy}, S.~D. and {Beardmore}, A.~P. and {Burrows}, D.~N. and {Pagani}, C. and {Starling}, R.~L.~C. and {Gehrels}, N. and {Romano}, P.},
        title = "{The Swift Burst Analyser. I. BAT and XRT spectral and flux evolution of gamma ray bursts}",
      journal = {\aap},
         year = 2010,
        month = sep,
       volume = {519},
          eid = {A102},
        pages = {A102},
          doi = {10.1051/0004-6361/201014819},
archivePrefix = {arXiv},
       eprint = {1004.3208},
 primaryClass = {astro-ph.IM},
       adsurl = {https://ui.adsabs.harvard.edu/abs/2010A&A...519A.102E}
}

@ARTICLE{Troja_22,
       author = {{Troja}, E. and {Fryer}, C.~L. and {O'Connor}, B. and {Ryan}, G. and {Dichiara}, S. and {Kumar}, A. and {Ito}, N. and {Gupta}, R. and {Wollaeger}, R.~T. and {Norris}, J.~P. and {Kawai}, N. and {Butler}, N.~R. and {Aryan}, A. and {Misra}, K. and {Hosokawa}, R. and {Murata}, K.~L. and {Niwano}, M. and {Pandey}, S.~B. and {Kutyrev}, A. and {van Eerten}, H.~J. and {Chase}, E.~A. and {Hu}, Y.-D. and {Caballero-Garcia}, M.~D. and {Castro-Tirado}, A.~J.},
        title = "{A nearby long gamma-ray burst from a merger of compact objects}",
      journal = {\nat},
         year = 2022,
        month = dec,
       volume = {612},
       number = {7939},
        pages = {228-231},
          doi = {10.1038/s41586-022-05327-3},
archivePrefix = {arXiv},
       eprint = {2209.03363},
 primaryClass = {astro-ph.HE},
       adsurl = {https://ui.adsabs.harvard.edu/abs/2022Natur.612..228T}
}

@ARTICLE{Rastinejad_22,
       author = {{Rastinejad}, Jillian C. and {Gompertz}, Benjamin P. and {Levan}, Andrew J. and {Fong}, Wen-fai and {Nicholl}, Matt and {Lamb}, Gavin P. and {Malesani}, Daniele B. and {Nugent}, Anya E. and {Oates}, Samantha R. and {Tanvir}, Nial R. and {de Ugarte Postigo}, Antonio and {Kilpatrick}, Charles D. and {Moore}, Christopher J. and {Metzger}, Brian D. and {Ravasio}, Maria Edvige and {Rossi}, Andrea and {Schroeder}, Genevieve and {Jencson}, Jacob and {Sand}, David J. and {Smith}, Nathan and {Ag{\"u}{\'\i} Fern{\'a}ndez}, Jos{\'e} Feliciano and {Berger}, Edo and {Blanchard}, Peter K. and {Chornock}, Ryan and {Cobb}, Bethany E. and {De Pasquale}, Massimiliano and {Fynbo}, Johan P.~U. and {Izzo}, Luca and {Kann}, D. Alexander and {Laskar}, Tanmoy and {Marini}, Ester and {Paterson}, Kerry and {Escorial}, Alicia Rouco and {Sears}, Huei M. and {Th{\"o}ne}, Christina C.},
        title = "{A kilonova following a long-duration gamma-ray burst at 350 Mpc}",
      journal = {\nat},
         year = 2022,
        month = dec,
       volume = {612},
       number = {7939},
        pages = {223-227},
          doi = {10.1038/s41586-022-05390-w},
archivePrefix = {arXiv},
       eprint = {2204.10864},
 primaryClass = {astro-ph.HE},
       adsurl = {https://ui.adsabs.harvard.edu/abs/2022Natur.612..223R}
}

@ARTICLE{Band_1997,
       author = {{Band}, David L.},
        title = "{Gamma-Ray Burst Spectral Evolution through Cross-Correlations of Discriminator Light Curves}",
      journal = {\apj},
         year = 1997,
        month = sep,
       volume = {486},
       number = {2},
        pages = {928-937},
          doi = {10.1086/304566},
archivePrefix = {arXiv},
       eprint = {astro-ph/9704206},
 primaryClass = {astro-ph},
       adsurl = {https://ui.adsabs.harvard.edu/abs/1997ApJ...486..928B}
}

@ARTICLE{Berger14,
       author = {{Berger}, Edo},
        title = "{Short-Duration Gamma-Ray Bursts}",
      journal = {\araa},
         year = 2014,
        month = aug,
       volume = {52},
        pages = {43-105},
          doi = {10.1146/annurev-astro-081913-035926},
archivePrefix = {arXiv},
       eprint = {1311.2603},
 primaryClass = {astro-ph.HE},
       adsurl = {https://ui.adsabs.harvard.edu/abs/2014ARA&A..52...43B}
}

@article{Bernardini_2015,
    author = {Bernardini, M. G. and Ghirlanda, G. and Campana, S. and Covino, S. and Salvaterra, R. and Atteia, J.-L. and Burlon, D. and Calderone, G. and D'Avanzo, P. and D'Elia, V. and Ghisellini, G. and Heussaff, V. and Lazzati, D. and Melandri, A. and Nava, L. and Vergani, S. D. and Tagliaferri, G.},
    title = {Comparing the spectral lag of short and long gamma-ray bursts and its relation with the luminosity},
    journal = {Monthly Notices of the Royal Astronomical Society},
    volume = {446},
    number = {2},
    pages = {1129-1138},
    year = {2015},
    month = {01},
    issn = {0035-8711},
    doi = {10.1093/mnras/stu2153},
    url = {https://doi.org/10.1093/mnras/stu2153},
    eprint = {https://academic.oup.com/mnras/article-pdf/446/2/1129/9380143/stu2153.pdf},
}

@ARTICLE{Bhat_2013,
       author = {{Bhat}, P.~N. and {Briggs}, Michael S. and {Connaughton}, Valerie and {Kouveliotou}, Chryssa and {van der Horst}, Alexander J. and {Paciesas}, William and {Meegan}, Charles A. and {Bissaldi}, Elisabetta and {Burgess}, Michael and {Chaplin}, Vandiver and {Diehl}, Roland and {Fishman}, Gerald and {Fitzpatrick}, Gerard and {Foley}, Suzanne and {Gibby}, Melissa and {Giles}, Misty M. and {Goldstein}, Adam and {Greiner}, Jochen and {Gruber}, David and {Guiriec}, Sylvain and {von Kienlin}, Andreas and {Kippen}, Marc and {McBreen}, Sheila and {Preece}, Robert and {Rau}, Arne and {Tierney}, Dave and {Wilson-Hodge}, Colleen},
        title = "{Temporal Deconvolution Study of Long and Short Gamma-Ray Burst Light Curves}",
      journal = {\apj},
         year = 2012,
        month = jan,
       volume = {744},
       number = {2},
          eid = {141},
        pages = {141},
          doi = {10.1088/0004-637X/744/2/141},
archivePrefix = {arXiv},
       eprint = {1109.4064},
 primaryClass = {astro-ph.HE},
       adsurl = {https://ui.adsabs.harvard.edu/abs/2012ApJ...744..141B}
}

@ARTICLE{cheng_95,
       author = {{Cheng}, L.~X. and {Ma}, Y.~Q. and {Cheng}, K.~S. and {Lu}, T. and {Zhou}, Y.~Y.},
        title = "{The time delay of gamma-ray bursts in the soft energy band.}",
      journal = {\aap},
         year = 1995,
        month = aug,
       volume = {300},
        pages = {746},
       adsurl = {https://ui.adsabs.harvard.edu/abs/1995A&A...300..746C}
}

@article{Cowperthwaite17,
doi = {10.3847/2041-8213/aa8fc7},
url = {https://doi.org/10.3847/2041-8213/aa8fc7},
year = {2017},
month = {oct},
publisher = {The American Astronomical Society},
volume = {848},
number = {2},
pages = {L17},
author = {Cowperthwaite, P. S. and Berger, E. and Villar, V. A. and Metzger, B. D. and Nicholl, M. and Chornock, R. and Blanchard, P. K. and Fong, W. and Margutti, R. and Soares-Santos, M. and Alexander, K. D. and Allam, S. and Annis, J. and Brout, D. and Brown, D. A. and Butler, R. E. and Chen, H.-Y. and Diehl, H. T. and Doctor, Z. and Drout, M. R. and Eftekhari, T. and Farr, B. and Finley, D. A. and Foley, R. J. and Frieman, J. A. and Fryer, C. L. and García-Bellido, J. and Gill, M. S. S. and Guillochon, J. and Herner, K. and Holz, D. E. and Kasen, D. and Kessler, R. and Marriner, J. and Matheson, T. and Neilsen, E. H. and Quataert, E. and Palmese, A. and Rest, A. and Sako, M. and Scolnic, D. M. and Smith, N. and Tucker, D. L. and Williams, P. K. G. and Balbinot, E. and Carlin, J. L. and Cook, E. R. and Durret, F. and Li, T. S. and Lopes, P. A. A. and Lourenço, A. C. C. and Marshall, J. L. and Medina, G. E. and Muir, J. and Muñoz, R. R. and Sauseda, M. and Schlegel, D. J. and Secco, L. F. and Vivas, A. K. and Wester, W. and Zenteno, A. and Zhang, Y. and Abbott, T. M. C. and Banerji, M. and Bechtol, K. and Benoit-Lévy, A. and Bertin, E. and Buckley-Geer, E. and Burke, D. L. and Capozzi, D. and Carnero Rosell, A. and Carrasco Kind, M. and Castander, F. J. and Crocce, M. and Cunha, C. E. and D’Andrea, C. B. and Costa, L. N. da and Davis, C. and DePoy, D. L. and Desai, S. and Dietrich, J. P. and Drlica-Wagner, A. and Eifler, T. F. and Evrard, A. E. and Fernandez, E. and Flaugher, B. and Fosalba, P. and Gaztanaga, E. and Gerdes, D. W. and Giannantonio, T. and Goldstein, D. A. and Gruen, D. and Gruendl, R. A. and Gutierrez, G. and Honscheid, K. and Jain, B. and James, D. J. and Jeltema, T. and Johnson, M. W. G. and Johnson, M. D. and Kent, S. and Krause, E. and Kron, R. and Kuehn, K. and Nuropatkin, N. and Lahav, O. and Lima, M. and Lin, H. and Maia, M. A. G. and March, M. and Martini, P. and McMahon, R. G. and Menanteau, F. and Miller, C. J. and Miquel, R. and Mohr, J. J. and Neilsen, E. and Nichol, R. C. and Ogando, R. L. C. and Plazas, A. A. and Roe, N. and Romer, A. K. and Roodman, A. and Rykoff, E. S. and Sanchez, E. and Scarpine, V. and Schindler, R. and Schubnell, M. and Sevilla-Noarbe, I. and Smith, M. and Smith, R. C. and Sobreira, F. and Suchyta, E. and Swanson, M. E. C. and Tarle, G. and Thomas, D. and Thomas, R. C. and Troxel, M. A. and Vikram, V. and Walker, A. R. and Wechsler, R. H. and Weller, J. and Yanny, B. and Zuntz, J.},
title = {The Electromagnetic Counterpart of the Binary Neutron Star Merger LIGO/Virgo GW170817. II. UV, Optical, and Near-infrared Light Curves and Comparison to Kilonova Models},
journal = {The Astrophysical Journal Letters}
}

@article{Davanzo14,
    author = {D'Avanzo, P. and Salvaterra, R. and Bernardini, M. G. and Nava, L. and Campana, S. and Covino, S. and D'Elia, V. and Ghirlanda, G. and Ghisellini, G. and Melandri, A. and Sbarufatti, B. and Vergani, S. D. and Tagliaferri, G.},
    title = {A complete sample of bright Swift short gamma-ray bursts},
    journal = {Monthly Notices of the Royal Astronomical Society},
    volume = {442},
    number = {3},
    pages = {2342-2356},
    year = {2014},
    month = {08},
    issn = {0035-8711},
    doi = {10.1093/mnras/stu994},
    url = {https://doi.org/10.1093/mnras/stu994},
    eprint = {https://academic.oup.com/mnras/article-pdf/442/3/2342/3599190/stu994.pdf},
}

@ARTICLE{Dermer_1998,
       author = {{Dermer}, Charles D.},
        title = "{On Spectral and Temporal Variability in Blazars and Gamma-Ray Bursts}",
      journal = {\apjl},
         year = 1998,
        month = jul,
       volume = {501},
       number = {2},
        pages = {L157-L161},
          doi = {10.1086/311467},
archivePrefix = {arXiv},
       eprint = {astro-ph/9805289},
 primaryClass = {astro-ph},
       adsurl = {https://ui.adsabs.harvard.edu/abs/1998ApJ...501L.157D}
}

@ARTICLE{Drenkhahn_2002,
       author = {{Drenkhahn}, G. and {Spruit}, H.~C.},
        title = "{Efficient acceleration and radiation in Poynting flux powered GRB outflows}",
      journal = {\aap},
         year = 2002,
        month = sep,
       volume = {391},
        pages = {1141-1153},
          doi = {10.1051/0004-6361:20020839},
archivePrefix = {arXiv},
       eprint = {astro-ph/0202387},
 primaryClass = {astro-ph},
       adsurl = {https://ui.adsabs.harvard.edu/abs/2002A&A...391.1141D}
}

@ARTICLE{Eichler89,
       author = {{Eichler}, David and {Livio}, Mario and {Piran}, Tsvi and {Schramm}, David N.},
        title = "{Nucleosynthesis, neutrino bursts and {\ensuremath{\gamma}}-rays from coalescing neutron stars}",
      journal = {\nat},
         year = 1989,
        month = jul,
       volume = {340},
       number = {6229},
        pages = {126-128},
          doi = {10.1038/340126a0},
       adsurl = {https://ui.adsabs.harvard.edu/abs/1989Natur.340..126E}
}

@ARTICLE{Godet_26,
       author = {{Godet}, O. and {Atteia}, J.-L. and {Schanne}, S. and {Lachaud}, C. and {Goldwurm}, A. and {Piron}, F. and {Guillemot}, Ph. and {Amoros}, C. and {Bertoli}, W. and {Bouchet}, L. and {Charmeau}, M.~C. and {Chteau}, F. and {Cordier}, B. and {Dagoneau}, N. and {Daly}, F. and {Dezalay}, J.-P. and {Galezzi}, J. and {Givaudan}, A. and {Gros}, A. and {Karakac}, M. and {Lacombe}, K. and {Leprovost}, H. and {Maestre}, S. and {Mercier}, K. and {Pasquier}, H. and {Perraud}, L. and {Pons}, R. and {Rambaud}, D. and {Simonella}, O. and {Tourrette}, T. and {Triou}, H. and {Waegebaert}, V. and {Bacon}, P. and {Barlyaeva}, T. and {Bellemont}, N. and {Bernardini}, M.-G. and {Brunet}, M. and {Cangemi}, F. and {Cavet}, C. and {Coleiro}, A. and {Corre}, D. and {Daigne}, F. and {Foisseau}, A. and {Gevin}, O. and {Guillot}, S. and {Jacob}, U. and {Lacreu}, F. and {Le Stum}, S. and {Maeght}, P. and {Maiolino}, T. and {Maolo}, A. and {Tcherniatinksy}, G. and {Wang}, J. and {Yang}, H.},
        title = "{ECLAIRs: the SVOM high-energy transient trigger camera}",
      journal = {arXiv e-prints},
         year = 2026,
        month = apr,
          eid = {arXiv:2604.24249},
        pages = {arXiv:2604.24249},
          doi = {10.48550/arXiv.2604.24249},
archivePrefix = {arXiv},
       eprint = {2604.24249},
 primaryClass = {astro-ph.HE},
       adsurl = {https://ui.adsabs.harvard.edu/abs/2026arXiv260424249G}
}

@ARTICLE{Wei_16,
       author = {{Wei}, J. and {Cordier}, B. and {Antier}, S. and {Antilogus}, P. and {Atteia}, J.-L. and {Bajat}, A. and {Basa}, S. and {Beckmann}, V. and {Bernardini}, M.~G. and {Boissier}, S. and {Bouchet}, L. and {Burwitz}, V. and {Claret}, A. and {Dai}, Z.-G. and {Daigne}, F. and {Deng}, J. and {Dornic}, D. and {Feng}, H. and {Foglizzo}, T. and {Gao}, H. and {Gehrels}, N. and {Godet}, O. and {Goldwurm}, A. and {Gonzalez}, F. and {Gosset}, L. and {G{\"o}tz}, D. and {Gouiffes}, C. and {Grise}, F. and {Gros}, A. and {Guilet}, J. and {Han}, X. and {Huang}, M. and {Huang}, Y.-F. and {Jouret}, M. and {Klotz}, A. and {La Marle}, O. and {Lachaud}, C. and {Le Floch}, E. and {Lee}, W. and {Leroy}, N. and {Li}, L.-X. and {Li}, S.~C. and {Li}, Z. and {Liang}, E.-W. and {Lyu}, H. and {Mercier}, K. and {Migliori}, G. and {Mochkovitch}, R. and {O'Brien}, P. and {Osborne}, J. and {Paul}, J. and {Perinati}, E. and {Petitjean}, P. and {Piron}, F. and {Qiu}, Y. and {Rau}, A. and {Rodriguez}, J. and {Schanne}, S. and {Tanvir}, N. and {Vangioni}, E. and {Vergani}, S. and {Wang}, F.-Y. and {Wang}, J. and {Wang}, X.-G. and {Wang}, X.-Y. and {Watson}, A. and {Webb}, N. and {Wei}, J.~J. and {Willingale}, R. and {Wu}, C. and {Wu}, X.-F. and {Xin}, L.-P. and {Xu}, D. and {Yu}, S. and {Yu}, W.-F. and {Yu}, Y.-W. and {Zhang}, B. and {Zhang}, S.-N. and {Zhang}, Y. and {Zhou}, X.~L.},
        title = "{The Deep and Transient Universe in the SVOM Era: New Challenges and Opportunities - Scientific prospects of the SVOM mission}",
      journal = {arXiv e-prints},
         year = 2016,
        month = oct,
          eid = {arXiv:1610.06892},
        pages = {arXiv:1610.06892},
          doi = {10.48550/arXiv.1610.06892},
archivePrefix = {arXiv},
       eprint = {1610.06892},
 primaryClass = {astro-ph.IM},
       adsurl = {https://ui.adsabs.harvard.edu/abs/2016arXiv161006892W}
}

@ARTICLE{Cordier_15,
       author = {{Cordier}, B. and {Wei}, J. and {Atteia}, J.-L. and {Basa}, S. and {Claret}, A. and {Daigne}, F. and {Deng}, J. and {Dong}, Y. and {Godet}, O. and {Goldwurm}, A. and {G{\"o}tz}, D. and {Han}, X. and {Klotz}, A. and {Lachaud}, C. and {Osborne}, J. and {Qiu}, Y. and {Schanne}, S. and {Wu}, B. and {Wang}, J. and {Wu}, C and {Xin}, L. and {Zhang}, B. and {Zhang}, S.-N.},
        title = "{The SVOM gamma-ray burst mission}",
      journal = {arXiv e-prints},
         year = 2015,
        month = dec,
          eid = {arXiv:1512.03323},
        pages = {arXiv:1512.03323},
          doi = {10.48550/arXiv.1512.03323},
archivePrefix = {arXiv},
       eprint = {1512.03323},
 primaryClass = {astro-ph.IM},
       adsurl = {https://ui.adsabs.harvard.edu/abs/2015arXiv151203323C}
}

@ARTICLE{Davanzo_12,
       author = {{D'Avanzo}, P. and {Salvaterra}, R. and {Sbarufatti}, B. and {Nava}, L. and {Melandri}, A. and {Bernardini}, M.~G. and {Campana}, S. and {Covino}, S. and {Fugazza}, D. and {Ghirlanda}, G. and {Ghisellini}, G. and {La Parola}, V. and {Perri}, M. and {Vergani}, S.~D. and {Tagliaferri}, G.},
        title = "{A complete sample of bright Swift Gamma-ray bursts: X-ray afterglow luminosity and its correlation with the prompt emission}",
      journal = {\mnras},
         year = 2012,
        month = sep,
       volume = {425},
       number = {1},
        pages = {506-513},
          doi = {10.1111/j.1365-2966.2012.21489.x},
archivePrefix = {arXiv},
       eprint = {1206.2357},
 primaryClass = {astro-ph.HE},
       adsurl = {https://ui.adsabs.harvard.edu/abs/2012MNRAS.425..506D}
}

@ARTICLE{evans17,
       author = {{Evans}, P.~A. and {Cenko}, S.~B. and {Kennea}, J.~A. and {Emery}, S.~W.~K. and {Kuin}, N.~P.~M. and {Korobkin}, O. and {Wollaeger}, R.~T. and {Fryer}, C.~L. and {Madsen}, K.~K. and {Harrison}, F.~A. and {Xu}, Y. and {Nakar}, E. and {Hotokezaka}, K. and {Lien}, A. and {Campana}, S. and {Oates}, S.~R. and {Troja}, E. and {Breeveld}, A.~A. and {Marshall}, F.~E. and {Barthelmy}, S.~D. and {Beardmore}, A.~P. and {Burrows}, D.~N. and {Cusumano}, G. and {D'A{\`\i}}, A. and {D'Avanzo}, P. and {D'Elia}, V. and {de Pasquale}, M. and {Even}, W.~P. and {Fontes}, C.~J. and {Forster}, K. and {Garcia}, J. and {Giommi}, P. and {Grefenstette}, B. and {Gronwall}, C. and {Hartmann}, D.~H. and {Heida}, M. and {Hungerford}, A.~L. and {Kasliwal}, M.~M. and {Krimm}, H.~A. and {Levan}, A.~J. and {Malesani}, D. and {Melandri}, A. and {Miyasaka}, H. and {Nousek}, J.~A. and {O'Brien}, P.~T. and {Osborne}, J.~P. and {Pagani}, C. and {Page}, K.~L. and {Palmer}, D.~M. and {Perri}, M. and {Pike}, S. and {Racusin}, J.~L. and {Rosswog}, S. and {Siegel}, M.~H. and {Sakamoto}, T. and {Sbarufatti}, B. and {Tagliaferri}, G. and {Tanvir}, N.~R. and {Tohuvavohu}, A.},
        title = "{Swift and NuSTAR observations of GW170817: Detection of a blue kilonova}",
      journal = {Science},
         year = 2017,
        month = dec,
       volume = {358},
       number = {6370},
        pages = {1565-1570},
          doi = {10.1126/science.aap9580},
archivePrefix = {arXiv},
       eprint = {1710.05437},
 primaryClass = {astro-ph.HE},
       adsurl = {https://ui.adsabs.harvard.edu/abs/2017Sci...358.1565E}
}

@ARTICLE{Ford_1995,
       author = {{Ford}, L.~A. and {Band}, D.~L. and {Matteson}, J.~L. and {Briggs}, M.~S. and {Pendleton}, G.~N. and {Preece}, R.~D. and {Paciesas}, W.~S. and {Teegarden}, B.~J. and {Palmer}, D.~M. and {Schaefer}, B.~E. and {Cline}, T.~L. and {Fishman}, G.~J. and {Kouveliotou}, C. and {Meegan}, C.~A. and {Wilson}, R.~B. and {Lestrade}, J.~P.},
        title = "{BATSE Observations of Gamma-Ray Burst Spectra. II. Peak Energy Evolution in Bright, Long Bursts}",
      journal = {\apj},
         year = 1995,
        month = jan,
       volume = {439},
        pages = {307},
          doi = {10.1086/175174},
archivePrefix = {arXiv},
       eprint = {astro-ph/9407090},
 primaryClass = {astro-ph},
       adsurl = {https://ui.adsabs.harvard.edu/abs/1995ApJ...439..307F}
}

@ARTICLE{Frail2001,
       author = {{Frail}, D.~A. and {Kulkarni}, S.~R. and {Sari}, R. and {Djorgovski}, S.~G. and {Bloom}, J.~S. and {Galama}, T.~J. and {Reichart}, D.~E. and {Berger}, E. and {Harrison}, F.~A. and {Price}, P.~A. and {Yost}, S.~A. and {Diercks}, A. and {Goodrich}, R.~W. and {Chaffee}, F.},
        title = "{Beaming in Gamma-Ray Bursts: Evidence for a Standard Energy Reservoir}",
      journal = {\apjl},
         year = 2001,
        month = nov,
       volume = {562},
       number = {1},
        pages = {L55-L58},
          doi = {10.1086/338119},
archivePrefix = {arXiv},
       eprint = {astro-ph/0102282},
 primaryClass = {astro-ph},
       adsurl = {https://ui.adsabs.harvard.edu/abs/2001ApJ...562L..55F}
}

@ARTICLE{Barthelmy_05,
       author = {{Barthelmy}, Scott D. and {Barbier}, Louis M. and {Cummings}, Jay R. and {Fenimore}, Ed E. and {Gehrels}, Neil and {Hullinger}, Derek and {Krimm}, Hans A. and {Markwardt}, Craig B. and {Palmer}, David M. and {Parsons}, Ann and {Sato}, Goro and {Suzuki}, Masaya and {Takahashi}, Tadayuki and {Tashiro}, Makota and {Tueller}, Jack},
        title = "{The Burst Alert Telescope (BAT) on the SWIFT Midex Mission}",
      journal = {\ssr},
         year = 2005,
        month = oct,
       volume = {120},
       number = {3-4},
        pages = {143-164},
          doi = {10.1007/s11214-005-5096-3},
archivePrefix = {arXiv},
       eprint = {astro-ph/0507410},
 primaryClass = {astro-ph},
       adsurl = {https://ui.adsabs.harvard.edu/abs/2005SSRv..120..143B}
}

@INPROCEEDINGS{Arnaud_96,
       author = {{Arnaud}, K.~A.},
        title = "{XSPEC: The First Ten Years}",
    booktitle = {Astronomical Data Analysis Software and Systems V},
         year = 1996,
       editor = {{Jacoby}, George H. and {Barnes}, Jeannette},
       series = {Astronomical Society of the Pacific Conference Series},
       volume = {101},
        month = jan,
        pages = {17},
       adsurl = {https://ui.adsabs.harvard.edu/abs/1996ASPC..101...17A}
}

@ARTICLE{Burrows_05,
       author = {{Burrows}, David N. and {Hill}, J.~E. and {Nousek}, J.~A. and {Kennea}, J.~A. and {Wells}, A. and {Osborne}, J.~P. and {Abbey}, A.~F. and {Beardmore}, A. and {Mukerjee}, K. and {Short}, A.~D.~T. and {Chincarini}, G. and {Campana}, S. and {Citterio}, O. and {Moretti}, A. and {Pagani}, C. and {Tagliaferri}, G. and {Giommi}, P. and {Capalbi}, M. and {Tamburelli}, F. and {Angelini}, L. and {Cusumano}, G. and {Br{\"a}uninger}, H.~W. and {Burkert}, W. and {Hartner}, G.~D.},
        title = "{The Swift X-Ray Telescope}",
      journal = {\ssr},
         year = 2005,
        month = oct,
       volume = {120},
       number = {3-4},
        pages = {165-195},
          doi = {10.1007/s11214-005-5097-2},
archivePrefix = {arXiv},
       eprint = {astro-ph/0508071},
 primaryClass = {astro-ph},
       adsurl = {https://ui.adsabs.harvard.edu/abs/2005SSRv..120..165B}
}

@ARTICLE{Meegan_09,
       author = {{Meegan}, Charles and {Lichti}, Giselher and {Bhat}, P.~N. and {Bissaldi}, Elisabetta and {Briggs}, Michael S. and {Connaughton}, Valerie and {Diehl}, Roland and {Fishman}, Gerald and {Greiner}, Jochen and {Hoover}, Andrew S. and {van der Horst}, Alexander J. and {von Kienlin}, Andreas and {Kippen}, R. Marc and {Kouveliotou}, Chryssa and {McBreen}, Sheila and {Paciesas}, W.~S. and {Preece}, Robert and {Steinle}, Helmut and {Wallace}, Mark S. and {Wilson}, Robert B. and {Wilson-Hodge}, Colleen},
        title = "{The Fermi Gamma-ray Burst Monitor}",
      journal = {\apj},
         year = 2009,
        month = sep,
       volume = {702},
       number = {1},
        pages = {791-804},
          doi = {10.1088/0004-637X/702/1/791},
archivePrefix = {arXiv},
       eprint = {0908.0450},
 primaryClass = {astro-ph.IM},
       adsurl = {https://ui.adsabs.harvard.edu/abs/2009ApJ...702..791M}
}

@ARTICLE{Perley_09,
       author = {{Perley}, D.~A. and {Metzger}, B.~D. and {Granot}, J. and {Butler}, N.~R. and {Sakamoto}, T. and {Ramirez-Ruiz}, E. and {Levan}, A.~J. and {Bloom}, J.~S. and {Miller}, A.~A. and {Bunker}, A. and {Chen}, H.-W. and {Filippenko}, A.~V. and {Gehrels}, N. and {Glazebrook}, K. and {Hall}, P.~B. and {Hurley}, K.~C. and {Kocevski}, D. and {Li}, W. and {Lopez}, S. and {Norris}, J. and {Piro}, A.~L. and {Poznanski}, D. and {Prochaska}, J.~X. and {Quataert}, E. and {Tanvir}, N.},
        title = "{GRB 080503: Implications of a Naked Short Gamma-Ray Burst Dominated by Extended Emission}",
      journal = {\apj},
         year = 2009,
        month = may,
       volume = {696},
       number = {2},
        pages = {1871-1885},
          doi = {10.1088/0004-637X/696/2/1871},
archivePrefix = {arXiv},
       eprint = {0811.1044},
 primaryClass = {astro-ph},
       adsurl = {https://ui.adsabs.harvard.edu/abs/2009ApJ...696.1871P}
}

@ARTICLE{Gehrels_04,
       author = {{Gehrels}, N. and {Chincarini}, G. and {Giommi}, P. and {Mason}, K.~O. and {Nousek}, J.~A. and {Wells}, A.~A. and {White}, N.~E. and {Barthelmy}, S.~D. and {Burrows}, D.~N. and {Cominsky}, L.~R. and {Hurley}, K.~C. and {Marshall}, F.~E. and {M{\'e}sz{\'a}ros}, P. and {Roming}, P.~W.~A. and {Angelini}, L. and {Barbier}, L.~M. and {Belloni}, T. and {Campana}, S. and {Caraveo}, P.~A. and {Chester}, M.~M. and {Citterio}, O. and {Cline}, T.~L. and {Cropper}, M.~S. and {Cummings}, J.~R. and {Dean}, A.~J. and {Feigelson}, E.~D. and {Fenimore}, E.~E. and {Frail}, D.~A. and {Fruchter}, A.~S. and {Garmire}, G.~P. and {Gendreau}, K. and {Ghisellini}, G. and {Greiner}, J. and {Hill}, J.~E. and {Hunsberger}, S.~D. and {Krimm}, H.~A. and {Kulkarni}, S.~R. and {Kumar}, P. and {Lebrun}, F. and {Lloyd-Ronning}, N.~M. and {Markwardt}, C.~B. and {Mattson}, B.~J. and {Mushotzky}, R.~F. and {Norris}, J.~P. and {Osborne}, J. and {Paczynski}, B. and {Palmer}, D.~M. and {Park}, H.-S. and {Parsons}, A.~M. and {Paul}, J. and {Rees}, M.~J. and {Reynolds}, C.~S. and {Rhoads}, J.~E. and {Sasseen}, T.~P. and {Schaefer}, B.~E. and {Short}, A.~T. and {Smale}, A.~P. and {Smith}, I.~A. and {Stella}, L. and {Tagliaferri}, G. and {Takahashi}, T. and {Tashiro}, M. and {Townsley}, L.~K. and {Tueller}, J. and {Turner}, M.~J.~L. and {Vietri}, M. and {Voges}, W. and {Ward}, M.~J. and {Willingale}, R. and {Zerbi}, F.~M. and {Zhang}, W.~W.},
        title = "{The Swift Gamma-Ray Burst Mission}",
      journal = {\apj},
         year = 2004,
        month = aug,
       volume = {611},
       number = {2},
        pages = {1005-1020},
          doi = {10.1086/422091},
archivePrefix = {arXiv},
       eprint = {astro-ph/0405233},
 primaryClass = {astro-ph},
       adsurl = {https://ui.adsabs.harvard.edu/abs/2004ApJ...611.1005G}
}

@ARTICLE{Gehrels_2006,
       author = {{Gehrels}, N. and {Norris}, J.~P. and {Barthelmy}, S.~D. and {Granot}, J. and {Kaneko}, Y. and {Kouveliotou}, C. and {Markwardt}, C.~B. and {M{\'e}sz{\'a}ros}, P. and {Nakar}, E. and {Nousek}, J.~A. and {O'Brien}, P.~T. and {Page}, M. and {Palmer}, D.~M. and {Parsons}, A.~M. and {Roming}, P.~W.~A. and {Sakamoto}, T. and {Sarazin}, C.~L. and {Schady}, P. and {Stamatikos}, M. and {Woosley}, S.~E.},
        title = "{A new {\ensuremath{\gamma}}-ray burst classification scheme from GRB060614}",
      journal = {\nat},
         year = 2006,
        month = dec,
       volume = {444},
       number = {7122},
        pages = {1044-1046},
          doi = {10.1038/nature05376},
archivePrefix = {arXiv},
       eprint = {astro-ph/0610635},
 primaryClass = {astro-ph},
       adsurl = {https://ui.adsabs.harvard.edu/abs/2006Natur.444.1044G}
}

@ARTICLE{ghirlanda_2009,
       author = {{Ghirlanda}, G. and {Nava}, L. and {Ghisellini}, G. and {Celotti}, A. and {Firmani}, C.},
        title = "{Short versus long gamma-ray bursts: spectra, energetics, and luminosities}",
      journal = {\aap},
         year = 2009,
        month = mar,
       volume = {496},
       number = {3},
        pages = {585-595},
          doi = {10.1051/0004-6361/200811209},
archivePrefix = {arXiv},
       eprint = {0902.0983},
 primaryClass = {astro-ph.HE},
       adsurl = {https://ui.adsabs.harvard.edu/abs/2009A&A...496..585G}
}

@ARTICLE{Goldstein17,
       author = {{Goldstein}, A. and {Veres}, P. and {Burns}, E. and {Briggs}, M.~S. and {Hamburg}, R. and {Kocevski}, D. and {Wilson-Hodge}, C.~A. and {Preece}, R.~D. and {Poolakkil}, S. and {Roberts}, O.~J. and {Hui}, C.~M. and {Connaughton}, V. and {Racusin}, J. and {von Kienlin}, A. and {Dal Canton}, T. and {Christensen}, N. and {Littenberg}, T. and {Siellez}, K. and {Blackburn}, L. and {Broida}, J. and {Bissaldi}, E. and {Cleveland}, W.~H. and {Gibby}, M.~H. and {Giles}, M.~M. and {Kippen}, R.~M. and {McBreen}, S. and {McEnery}, J. and {Meegan}, C.~A. and {Paciesas}, W.~S. and {Stanbro}, M.},
        title = "{An Ordinary Short Gamma-Ray Burst with Extraordinary Implications: Fermi-GBM Detection of GRB 170817A}",
      journal = {\apjl},
         year = 2017,
        month = oct,
       volume = {848},
       number = {2},
          eid = {L14},
        pages = {L14},
          doi = {10.3847/2041-8213/aa8f41},
archivePrefix = {arXiv},
       eprint = {1710.05446},
 primaryClass = {astro-ph.HE},
       adsurl = {https://ui.adsabs.harvard.edu/abs/2017ApJ...848L..14G}
}

@ARTICLE{Gompertz_2013,
       author = {{Gompertz}, B.~P. and {O'Brien}, P.~T. and {Wynn}, G.~A.},
        title = "{Magnetar powered GRBs: explaining the extended emission and X-ray plateau of short GRB light curves}",
      journal = {\mnras},
         year = 2014,
        month = feb,
       volume = {438},
       number = {1},
        pages = {240-250},
          doi = {10.1093/mnras/stt2165},
archivePrefix = {arXiv},
       eprint = {1311.1505},
 primaryClass = {astro-ph.HE},
       adsurl = {https://ui.adsabs.harvard.edu/abs/2014MNRAS.438..240G}
}

@ARTICLE{Kaneko_2015,
       author = {{Kaneko}, Y. and {Bostanc{\i}}, Z.~F. and {G{\"o}{\u{g}}{\"u}{\textcommabelow s}}, E. and {Lin}, L.},
        title = "{Short gamma-ray bursts with extended emission observed with Swift/BAT and Fermi/GBM}",
      journal = {\mnras},
         year = 2015,
        month = sep,
       volume = {452},
       number = {1},
        pages = {824-837},
          doi = {10.1093/mnras/stv1286},
archivePrefix = {arXiv},
       eprint = {1506.05899},
 primaryClass = {astro-ph.HE},
       adsurl = {https://ui.adsabs.harvard.edu/abs/2015MNRAS.452..824K}
}

@ARTICLE{Kouveliotou93,
       author = {{Kouveliotou}, Chryssa and {Meegan}, Charles A. and {Fishman}, Gerald J. and {Bhat}, Narayana P. and {Briggs}, Michael S. and {Koshut}, Thomas M. and {Paciesas}, William S. and {Pendleton}, Geoffrey N.},
        title = "{Identification of Two Classes of Gamma-Ray Bursts}",
      journal = {\apjl},
         year = 1993,
        month = aug,
       volume = {413},
        pages = {L101},
          doi = {10.1086/186969},
       adsurl = {https://ui.adsabs.harvard.edu/abs/1993ApJ...413L.101K}
}

@ARTICLE{Levan_24,
       author = {{Levan}, Andrew J. and {Gompertz}, Benjamin P. and {Salafia}, Om Sharan and {Bulla}, Mattia and {Burns}, Eric and {Hotokezaka}, Kenta and {Izzo}, Luca and {Lamb}, Gavin P. and {Malesani}, Daniele B. and {Oates}, Samantha R. and {Ravasio}, Maria Edvige and {Rouco Escorial}, Alicia and {Schneider}, Benjamin and {Sarin}, Nikhil and {Schulze}, Steve and {Tanvir}, Nial R. and {Ackley}, Kendall and {Anderson}, Gemma and {Brammer}, Gabriel B. and {Christensen}, Lise and {Dhillon}, Vikram S. and {Evans}, Phil A. and {Fausnaugh}, Michael and {Fong}, Wen-fai and {Fruchter}, Andrew S. and {Fryer}, Chris and {Fynbo}, Johan P.~U. and {Gaspari}, Nicola and {Heintz}, Kasper E. and {Hjorth}, Jens and {Kennea}, Jamie A. and {Kennedy}, Mark R. and {Laskar}, Tanmoy and {Leloudas}, Giorgos and {Mandel}, Ilya and {Martin-Carrillo}, Antonio and {Metzger}, Brian D. and {Nicholl}, Matt and {Nugent}, Anya and {Palmerio}, Jesse T. and {Pugliese}, Giovanna and {Rastinejad}, Jillian and {Rhodes}, Lauren and {Rossi}, Andrea and {Saccardi}, Andrea and {Smartt}, Stephen J. and {Stevance}, Heloise F. and {Tohuvavohu}, Aaron and {van der Horst}, Alexander and {Vergani}, Susanna D. and {Watson}, Darach and {Barclay}, Thomas and {Bhirombhakdi}, Kornpob and {Breedt}, Elm{\'e} and {Breeveld}, Alice A. and {Brown}, Alexander J. and {Campana}, Sergio and {Chrimes}, Ashley A. and {D'Avanzo}, Paolo and {D'Elia}, Valerio and {De Pasquale}, Massimiliano and {Dyer}, Martin J. and {Galloway}, Duncan K. and {Garbutt}, James A. and {Green}, Matthew J. and {Hartmann}, Dieter H. and {Jakobsson}, P{\'a}ll and {Kerry}, Paul and {Kouveliotou}, Chryssa and {Langeroodi}, Danial and {Le Floc'h}, Emeric and {Leung}, James K. and {Littlefair}, Stuart P. and {Munday}, James and {O'Brien}, Paul and {Parsons}, Steven G. and {Pelisoli}, Ingrid and {Sahman}, David I. and {Salvaterra}, Ruben and {Sbarufatti}, Boris and {Steeghs}, Danny and {Tagliaferri}, Gianpiero and {Th{\"o}ne}, Christina C. and {de Ugarte Postigo}, Antonio and {Kann}, David Alexander},
        title = "{Heavy-element production in a compact object merger observed by JWST}",
      journal = {\nat},
         year = 2024,
        month = feb,
       volume = {626},
       number = {8000},
        pages = {737-741},
          doi = {10.1038/s41586-023-06759-1},
archivePrefix = {arXiv},
       eprint = {2307.02098},
 primaryClass = {astro-ph.HE},
       adsurl = {https://ui.adsabs.harvard.edu/abs/2024Natur.626..737L}
}

@ARTICLE{Rossi_22,
       author = {{Rossi}, A. and {Rothberg}, B. and {Palazzi}, E. and {Kann}, D.~A. and {D'Avanzo}, P. and {Amati}, L. and {Klose}, S. and {Perego}, A. and {Pian}, E. and {Guidorzi}, C. and {Pozanenko}, A.~S. and {Savaglio}, S. and {Stratta}, G. and {Agapito}, G. and {Covino}, S. and {Cusano}, F. and {D'Elia}, V. and {De Pasquale}, M. and {Della Valle}, M. and {Kuhn}, O. and {Izzo}, L. and {Loffredo}, E. and {Masetti}, N. and {Melandri}, A. and {Minaev}, P.~Y. and {Guelbenzu}, A. Nicuesa and {Paris}, D. and {Paiano}, S. and {Plantet}, C. and {Rossi}, F. and {Salvaterra}, R. and {Schulze}, S. and {Veillet}, C. and {Volnova}, A.~A.},
        title = "{The Peculiar Short-duration GRB 200826A and Its Supernova}",
      journal = {\apj},
         year = 2022,
        month = jun,
       volume = {932},
       number = {1},
          eid = {1},
        pages = {1},
          doi = {10.3847/1538-4357/ac60a2},
archivePrefix = {arXiv},
       eprint = {2105.03829},
 primaryClass = {astro-ph.HE},
       adsurl = {https://ui.adsabs.harvard.edu/abs/2022ApJ...932....1R}
}

@ARTICLE{Lien_17,
       author = {{Lien}, Amy and {Sakamoto}, Takanori and {Barthelmy}, Scott D. and {Baumgartner}, Wayne H. and {Cannizzo}, John K. and {Chen}, Kevin and {Collins}, Nicholas R. and {Cummings}, Jay R. and {Gehrels}, Neil and {Krimm}, Hans A. and {Markwardt}, Craig. B. and {Palmer}, David M. and {Stamatikos}, Michael and {Troja}, Eleonora and {Ukwatta}, T.~N.},
        title = "{The Third Swift Burst Alert Telescope Gamma-Ray Burst Catalog}",
      journal = {\apj},
         year = 2016,
        month = sep,
       volume = {829},
       number = {1},
          eid = {7},
        pages = {7},
          doi = {10.3847/0004-637X/829/1/7},
archivePrefix = {arXiv},
       eprint = {1606.01956},
 primaryClass = {astro-ph.HE},
       adsurl = {https://ui.adsabs.harvard.edu/abs/2016ApJ...829....7L}
}

@ARTICLE{Lysenko2025,
       author = {{Lysenko}, Alexandra L. and {Svinkin}, Dmitry S. and {Frederiks}, Dmitry D. and {Ridnaia}, Anna V. and {Tsvetkova}, Anastasia E. and {Ulanov}, Mikhail V.},
        title = "{The third Konus-Wind catalogue of short gamma-ray bursts}",
      journal = {\pasa},
         year = 2025,
        month = jun,
       volume = {42},
          eid = {e063},
        pages = {e063},
          doi = {10.1017/pasa.2025.10027},
archivePrefix = {arXiv},
       eprint = {2410.16896},
 primaryClass = {astro-ph.HE},
       adsurl = {https://ui.adsabs.harvard.edu/abs/2025PASA...42...63L}
}

@ARTICLE{Lu_2006,
       author = {{Lu}, R.-J. and {Qin}, Y.-P. and {Zhang}, Z.-B. and {Yi}, T.-F.},
        title = "{Spectral lags caused by the curvature effect of fireballs}",
      journal = {\mnras},
         year = 2006,
        month = mar,
       volume = {367},
       number = {1},
        pages = {275-289},
          doi = {10.1111/j.1365-2966.2005.09951.x},
archivePrefix = {arXiv},
       eprint = {astro-ph/0509287},
 primaryClass = {astro-ph},
       adsurl = {https://ui.adsabs.harvard.edu/abs/2006MNRAS.367..275L}
}

@ARTICLE{Lu_2022,
       author = {{L{\"u}}, Hou-Jun and {Yuan}, Hao-Yu and {Yi}, Ting-Feng and {Wang}, Xiang-Gao and {Hu}, You-Dong and {Yuan}, Yong and {Rice}, Jared and {Wang}, Jian-Guo and {Cao}, Jia-Xin and {Kong}, De-Feng and {Fernandez-Garc{\'\i}a}, Emilio and {Castro-Tirado}, Alberto J. and {Lian}, Ji-Shun and {Gan}, Wen-Pei and {Wang}, Shan-Qin and {Xin}, Li-Ping and {Caballero-Garc{\'\i}a}, M.~D. and {Fan}, Yu-Feng and {Liang}, En-Wei},
        title = "{GRB 211227A as a Peculiar Long Gamma-Ray Burst from a Compact Star Merger}",
      journal = {\apjl},
         year = 2022,
        month = jun,
       volume = {931},
       number = {2},
          eid = {L23},
        pages = {L23},
          doi = {10.3847/2041-8213/ac6e3a},
archivePrefix = {arXiv},
       eprint = {2201.06395},
 primaryClass = {astro-ph.HE},
       adsurl = {https://ui.adsabs.harvard.edu/abs/2022ApJ...931L..23L}
}

@ARTICLE{MacFadyen_1999,
       author = {{MacFadyen}, A.~I. and {Woosley}, S.~E.},
        title = "{Collapsars: Gamma-Ray Bursts and Explosions in ``Failed Supernovae''}",
      journal = {\apj},
         year = 1999,
        month = oct,
       volume = {524},
       number = {1},
        pages = {262-289},
          doi = {10.1086/307790},
archivePrefix = {arXiv},
       eprint = {astro-ph/9810274},
 primaryClass = {astro-ph},
       adsurl = {https://ui.adsabs.harvard.edu/abs/1999ApJ...524..262M}
}

@article{margutti2012,
    author = {Margutti, R. and Zaninoni, E. and Bernardini, M. G. and Chincarini, G. and Pasotti, F. and Guidorzi, C. and Angelini, L. and Burrows, D. N. and Capalbi, M. and Evans, P. A. and Gehrels, N. and Kennea, J. and Mangano, V. and Moretti, A. and Nousek, J. and Osborne, J. P. and Page, K. L. and Perri, M. and Racusin, J. and Romano, P. and Sbarufatti, B. and Stafford, S. and Stamatikos, M.},
    title = {The prompt-afterglow connection in gamma-ray bursts: a comprehensive statistical analysis of Swift X-ray light curves},
    journal = {Monthly Notices of the Royal Astronomical Society},
    volume = {428},
    number = {1},
    pages = {729-742},
    year = {2013},
    month = {01},
    issn = {0035-8711},
    doi = {10.1093/mnras/sts066},
    url = {https://doi.org/10.1093/mnras/sts066},
    eprint = {https://academic.oup.com/mnras/article-pdf/428/1/729/17335970/729.pdf},
}

@article{Meegan_2009,
doi = {10.1088/0004-637X/702/1/791},
url = {https://doi.org/10.1088/0004-637X/702/1/791},
year = {2009},
month = {aug},
publisher = {The American Astronomical Society},
volume = {702},
number = {1},
pages = {791},
author = {Meegan, Charles and Lichti, Giselher and Bhat, P. N. and Bissaldi, Elisabetta and Briggs, Michael S. and Connaughton, Valerie and Diehl, Roland and Fishman, Gerald and Greiner, Jochen and Hoover, Andrew S. and van der Horst, Alexander J. and von Kienlin, Andreas and Kippen, R. Marc and Kouveliotou, Chryssa and McBreen, Sheila and Paciesas, W. S. and Preece, Robert and Steinle, Helmut and Wallace, Mark S. and Wilson, Robert B. and Wilson-Hodge, Colleen},
title = {THE FERMI GAMMA-RAY BURST MONITOR},
journal = {The Astrophysical Journal}
}

@ARTICLE{Meszaros_1997,
       author = {{M{\'e}sz{\'a}ros}, P. and {Rees}, M.~J.},
        title = "{Optical and Long-Wavelength Afterglow from Gamma-Ray Bursts}",
      journal = {\apj},
         year = 1997,
        month = feb,
       volume = {476},
       number = {1},
        pages = {232-237},
          doi = {10.1086/303625},
archivePrefix = {arXiv},
       eprint = {astro-ph/9606043},
 primaryClass = {astro-ph},
       adsurl = {https://ui.adsabs.harvard.edu/abs/1997ApJ...476..232M}
}

@ARTICLE{Metzger_2008,
       author = {{Metzger}, B.~D. and {Quataert}, E. and {Thompson}, T.~A.},
        title = "{Short-duration gamma-ray bursts with extended emission from protomagnetar spin-down}",
      journal = {\mnras},
         year = 2008,
        month = apr,
       volume = {385},
       number = {3},
        pages = {1455-1460},
          doi = {10.1111/j.1365-2966.2008.12923.x},
archivePrefix = {arXiv},
       eprint = {0712.1233},
 primaryClass = {astro-ph},
       adsurl = {https://ui.adsabs.harvard.edu/abs/2008MNRAS.385.1455M}
}

@ARTICLE{Narayan92,
       author = {{Narayan}, Ramesh and {Paczynski}, Bohdan and {Piran}, Tsvi},
        title = "{Gamma-Ray Bursts as the Death Throes of Massive Binary Stars}",
      journal = {\apjl},
         year = 1992,
        month = aug,
       volume = {395},
        pages = {L83},
          doi = {10.1086/186493},
archivePrefix = {arXiv},
       eprint = {astro-ph/9204001},
 primaryClass = {astro-ph},
       adsurl = {https://ui.adsabs.harvard.edu/abs/1992ApJ...395L..83N}
}

@ARTICLE{Nava_2012,
       author = {{Nava}, L. and {Salvaterra}, R. and {Ghirlanda}, G. and {Ghisellini}, G. and {Campana}, S. and {Covino}, S. and {Cusumano}, G. and {D'Avanzo}, P. and {D'Elia}, V. and {Fugazza}, D. and {Melandri}, A. and {Sbarufatti}, B. and {Vergani}, S.~D. and {Tagliaferri}, G.},
        title = "{A complete sample of bright Swift long gamma-ray bursts: testing the spectral-energy correlations}",
      journal = {\mnras},
         year = 2012,
        month = apr,
       volume = {421},
       number = {2},
        pages = {1256-1264},
          doi = {10.1111/j.1365-2966.2011.20394.x},
archivePrefix = {arXiv},
       eprint = {1112.4470},
 primaryClass = {astro-ph.HE},
       adsurl = {https://ui.adsabs.harvard.edu/abs/2012MNRAS.421.1256N}
}

@ARTICLE{Norris_2000,
       author = {{Norris}, J.~P. and {Marani}, G.~F. and {Bonnell}, J.~T.},
        title = "{Connection between Energy-dependent Lags and Peak Luminosity in Gamma-Ray Bursts}",
      journal = {\apj},
         year = 2000,
        month = may,
       volume = {534},
       number = {1},
        pages = {248-257},
          doi = {10.1086/308725},
archivePrefix = {arXiv},
       eprint = {astro-ph/9903233},
 primaryClass = {astro-ph},
       adsurl = {https://ui.adsabs.harvard.edu/abs/2000ApJ...534..248N}
}

@ARTICLE{Norris_2006,
       author = {{Norris}, J.~P. and {Bonnell}, J.~T.},
        title = "{Short Gamma-Ray Bursts with Extended Emission}",
      journal = {\apj},
         year = 2006,
        month = may,
       volume = {643},
       number = {1},
        pages = {266-275},
          doi = {10.1086/502796},
archivePrefix = {arXiv},
       eprint = {astro-ph/0601190},
 primaryClass = {astro-ph},
       adsurl = {https://ui.adsabs.harvard.edu/abs/2006ApJ...643..266N}
}

@ARTICLE{Norris11,
       author = {{Norris}, Jay P. and {Gehrels}, Neil and {Scargle}, Jeffrey D.},
        title = "{Heterogeneity in Short Gamma-Ray Bursts}",
      journal = {\apj},
         year = 2011,
        month = jul,
       volume = {735},
       number = {1},
          eid = {23},
        pages = {23},
          doi = {10.1088/0004-637X/735/1/23},
archivePrefix = {arXiv},
       eprint = {1101.1648},
 primaryClass = {astro-ph.HE},
       adsurl = {https://ui.adsabs.harvard.edu/abs/2011ApJ...735...23N}
}

@ARTICLE{piran_04,
       author = {{Piran}, Tsvi},
        title = "{The physics of gamma-ray bursts}",
      journal = {Reviews of Modern Physics},
         year = 2004,
        month = oct,
       volume = {76},
       number = {4},
        pages = {1143-1210},
          doi = {10.1103/RevModPhys.76.1143},
archivePrefix = {arXiv},
       eprint = {astro-ph/0405503},
 primaryClass = {astro-ph},
       adsurl = {https://ui.adsabs.harvard.edu/abs/2004RvMP...76.1143P}
}

@ARTICLE{planck_18,
       author = {{Planck Collaboration} and {Aghanim}, N. and {Akrami}, Y. and {Ashdown}, M. and {Aumont}, J. and {Baccigalupi}, C. and {Ballardini}, M. and {Banday}, A.~J. and {Barreiro}, R.~B. and {Bartolo}, N. and {Basak}, S. and {Battye}, R. and {Benabed}, K. and {Bernard}, J.-P. and {Bersanelli}, M. and {Bielewicz}, P. and {Bock}, J.~J. and {Bond}, J.~R. and {Borrill}, J. and {Bouchet}, F.~R. and {Boulanger}, F. and {Bucher}, M. and {Burigana}, C. and {Butler}, R.~C. and {Calabrese}, E. and {Cardoso}, J.-F. and {Carron}, J. and {Challinor}, A. and {Chiang}, H.~C. and {Chluba}, J. and {Colombo}, L.~P.~L. and {Combet}, C. and {Contreras}, D. and {Crill}, B.~P. and {Cuttaia}, F. and {de Bernardis}, P. and {de Zotti}, G. and {Delabrouille}, J. and {Delouis}, J.-M. and {Di Valentino}, E. and {Diego}, J.~M. and {Dor{\'e}}, O. and {Douspis}, M. and {Ducout}, A. and {Dupac}, X. and {Dusini}, S. and {Efstathiou}, G. and {Elsner}, F. and {En{\ss}lin}, T.~A. and {Eriksen}, H.~K. and {Fantaye}, Y. and {Farhang}, M. and {Fergusson}, J. and {Fernandez-Cobos}, R. and {Finelli}, F. and {Forastieri}, F. and {Frailis}, M. and {Fraisse}, A.~A. and {Franceschi}, E. and {Frolov}, A. and {Galeotta}, S. and {Galli}, S. and {Ganga}, K. and {G{\'e}nova-Santos}, R.~T. and {Gerbino}, M. and {Ghosh}, T. and {Gonz{\'a}lez-Nuevo}, J. and {G{\'o}rski}, K.~M. and {Gratton}, S. and {Gruppuso}, A. and {Gudmundsson}, J.~E. and {Hamann}, J. and {Handley}, W. and {Hansen}, F.~K. and {Herranz}, D. and {Hildebrandt}, S.~R. and {Hivon}, E. and {Huang}, Z. and {Jaffe}, A.~H. and {Jones}, W.~C. and {Karakci}, A. and {Keih{\"a}nen}, E. and {Keskitalo}, R. and {Kiiveri}, K. and {Kim}, J. and {Kisner}, T.~S. and {Knox}, L. and {Krachmalnicoff}, N. and {Kunz}, M. and {Kurki-Suonio}, H. and {Lagache}, G. and {Lamarre}, J.-M. and {Lasenby}, A. and {Lattanzi}, M. and {Lawrence}, C.~R. and {Le Jeune}, M. and {Lemos}, P. and {Lesgourgues}, J. and {Levrier}, F. and {Lewis}, A. and {Liguori}, M. and {Lilje}, P.~B. and {Lilley}, M. and {Lindholm}, V. and {L{\'o}pez-Caniego}, M. and {Lubin}, P.~M. and {Ma}, Y.-Z. and {Mac{\'\i}as-P{\'e}rez}, J.~F. and {Maggio}, G. and {Maino}, D. and {Mandolesi}, N. and {Mangilli}, A. and {Marcos-Caballero}, A. and {Maris}, M. and {Martin}, P.~G. and {Martinelli}, M. and {Mart{\'\i}nez-Gonz{\'a}lez}, E. and {Matarrese}, S. and {Mauri}, N. and {McEwen}, J.~D. and {Meinhold}, P.~R. and {Melchiorri}, A. and {Mennella}, A. and {Migliaccio}, M. and {Millea}, M. and {Mitra}, S. and {Miville-Desch{\^e}nes}, M.-A. and {Molinari}, D. and {Montier}, L. and {Morgante}, G. and {Moss}, A. and {Natoli}, P. and {N{\o}rgaard-Nielsen}, H.~U. and {Pagano}, L. and {Paoletti}, D. and {Partridge}, B. and {Patanchon}, G. and {Peiris}, H.~V. and {Perrotta}, F. and {Pettorino}, V. and {Piacentini}, F. and {Polastri}, L. and {Polenta}, G. and {Puget}, J.-L. and {Rachen}, J.~P. and {Reinecke}, M. and {Remazeilles}, M. and {Renzi}, A. and {Rocha}, G. and {Rosset}, C. and {Roudier}, G. and {Rubi{\~n}o-Mart{\'\i}n}, J.~A. and {Ruiz-Granados}, B. and {Salvati}, L. and {Sandri}, M. and {Savelainen}, M. and {Scott}, D. and {Shellard}, E.~P.~S. and {Sirignano}, C. and {Sirri}, G. and {Spencer}, L.~D. and {Sunyaev}, R. and {Suur-Uski}, A.-S. and {Tauber}, J.~A. and {Tavagnacco}, D. and {Tenti}, M. and {Toffolatti}, L. and {Tomasi}, M. and {Trombetti}, T. and {Valenziano}, L. and {Valiviita}, J. and {Van Tent}, B. and {Vibert}, L. and {Vielva}, P. and {Villa}, F. and {Vittorio}, N. and {Wandelt}, B.~D. and {Wehus}, I.~K. and {White}, M. and {White}, S.~D.~M. and {Zacchei}, A. and {Zonca}, A.},
        title = "{Planck 2018 results. VI. Cosmological parameters}",
      journal = {\aap},
         year = 2020,
        month = sep,
       volume = {641},
          eid = {A6},
        pages = {A6},
          doi = {10.1051/0004-6361/201833910},
archivePrefix = {arXiv},
       eprint = {1807.06209},
 primaryClass = {astro-ph.CO},
       adsurl = {https://ui.adsabs.harvard.edu/abs/2020A&A...641A...6P}
}

@ARTICLE{Rees_1994,
       author = {{Rees}, M.~J. and {Meszaros}, P.},
        title = "{Unsteady Outflow Models for Cosmological Gamma-Ray Bursts}",
      journal = {\apjl},
         year = 1994,
        month = aug,
       volume = {430},
        pages = {L93},
          doi = {10.1086/187446},
archivePrefix = {arXiv},
       eprint = {astro-ph/9404038},
 primaryClass = {astro-ph},
       adsurl = {https://ui.adsabs.harvard.edu/abs/1994ApJ...430L..93R}
}

@ARTICLE{Rees_2005,
       author = {{Rees}, M.~J. and {M{\'e}sz{\'a}ros}, P.},
        title = "{Dissipative Photosphere Models of Gamma-Ray Bursts and X-Ray Flashes}",
      journal = {\apj},
         year = 2005,
        month = aug,
       volume = {628},
       number = {2},
        pages = {847-852},
          doi = {10.1086/430818},
archivePrefix = {arXiv},
       eprint = {astro-ph/0412702},
 primaryClass = {astro-ph},
       adsurl = {https://ui.adsabs.harvard.edu/abs/2005ApJ...628..847R}
}

@ARTICLE{Ryde_2005,
       author = {{Ryde}, F.},
        title = "{Interpretations of gamma-ray burst spectroscopy. I. Analytical and numerical study of spectral lags}",
      journal = {\aap},
         year = 2005,
        month = jan,
       volume = {429},
        pages = {869-879},
          doi = {10.1051/0004-6361:20041364},
archivePrefix = {arXiv},
       eprint = {astro-ph/0411206},
 primaryClass = {astro-ph},
       adsurl = {https://ui.adsabs.harvard.edu/abs/2005A&A...429..869R}
}

@ARTICLE{Rhoads1999,
       author = {{Rhoads}, James E.},
        title = "{The Dynamics and Light Curves of Beamed Gamma-Ray Burst Afterglows}",
      journal = {\apj},
         year = 1999,
        month = nov,
       volume = {525},
       number = {2},
        pages = {737-749},
          doi = {10.1086/307907},
archivePrefix = {arXiv},
       eprint = {astro-ph/9903399},
 primaryClass = {astro-ph},
       adsurl = {https://ui.adsabs.harvard.edu/abs/1999ApJ...525..737R}
}

@ARTICLE{Salmonson_2000,
       author = {{Salmonson}, Jay D.},
        title = "{On the Kinematic Origin of the Luminosity-Pulse Lag Relationship in Gamma-Ray Bursts}",
      journal = {\apjl},
         year = 2000,
        month = dec,
       volume = {544},
       number = {2},
        pages = {L115-L117},
          doi = {10.1086/317305},
archivePrefix = {arXiv},
       eprint = {astro-ph/0005264},
 primaryClass = {astro-ph},
       adsurl = {https://ui.adsabs.harvard.edu/abs/2000ApJ...544L.115S}
}

@ARTICLE{Salvaterra12,
       author = {{Salvaterra}, R. and {Campana}, S. and {Vergani}, S.~D. and {Covino}, S. and {D'Avanzo}, P. and {Fugazza}, D. and {Ghirlanda}, G. and {Ghisellini}, G. and {Melandri}, A. and {Nava}, L. and {Sbarufatti}, B. and {Flores}, H. and {Piranomonte}, S. and {Tagliaferri}, G.},
        title = "{A Complete Sample of Bright Swift Long Gamma-Ray Bursts. I. Sample Presentation, Luminosity Function and Evolution}",
      journal = {\apj},
         year = 2012,
        month = apr,
       volume = {749},
       number = {1},
          eid = {68},
        pages = {68},
          doi = {10.1088/0004-637X/749/1/68},
archivePrefix = {arXiv},
       eprint = {1112.1700},
 primaryClass = {astro-ph.CO},
       adsurl = {https://ui.adsabs.harvard.edu/abs/2012ApJ...749...68S}
}

@ARTICLE{Sari1998,
       author = {{Sari}, Re'em and {Piran}, Tsvi and {Narayan}, Ramesh},
        title = "{Spectra and Light Curves of Gamma-Ray Burst Afterglows}",
      journal = {\apjl},
         year = 1998,
        month = apr,
       volume = {497},
       number = {1},
        pages = {L17-L20},
          doi = {10.1086/311269},
archivePrefix = {arXiv},
       eprint = {astro-ph/9712005},
 primaryClass = {astro-ph},
       adsurl = {https://ui.adsabs.harvard.edu/abs/1998ApJ...497L..17S}
}

@ARTICLE{Sari1999,
       author = {{Sari}, Re'em and {Piran}, Tsvi and {Halpern}, J.~P.},
        title = "{Jets in Gamma-Ray Bursts}",
      journal = {\apjl},
         year = 1999,
        month = jul,
       volume = {519},
       number = {1},
        pages = {L17-L20},
          doi = {10.1086/312109},
archivePrefix = {arXiv},
       eprint = {astro-ph/9903339},
 primaryClass = {astro-ph},
       adsurl = {https://ui.adsabs.harvard.edu/abs/1999ApJ...519L..17S}
}

@ARTICLE{Savchenko17,
       author = {{Savchenko}, V. and {Ferrigno}, C. and {Kuulkers}, E. and {Bazzano}, A. and {Bozzo}, E. and {Brandt}, S. and {Chenevez}, J. and {Courvoisier}, T.~J.-L. and {Diehl}, R. and {Domingo}, A. and {Hanlon}, L. and {Jourdain}, E. and {von Kienlin}, A. and {Laurent}, P. and {Lebrun}, F. and {Lutovinov}, A. and {Martin-Carrillo}, A. and {Mereghetti}, S. and {Natalucci}, L. and {Rodi}, J. and {Roques}, J.-P. and {Sunyaev}, R. and {Ubertini}, P.},
        title = "{INTEGRAL Detection of the First Prompt Gamma-Ray Signal Coincident with the Gravitational-wave Event GW170817}",
      journal = {\apjl},
         year = 2017,
        month = oct,
       volume = {848},
       number = {2},
          eid = {L15},
        pages = {L15},
          doi = {10.3847/2041-8213/aa8f94},
archivePrefix = {arXiv},
       eprint = {1710.05449},
 primaryClass = {astro-ph.HE},
       adsurl = {https://ui.adsabs.harvard.edu/abs/2017ApJ...848L..15S}
}

@ARTICLE{Svinkin2016,
       author = {{Svinkin}, D.~S. and {Frederiks}, D.~D. and {Aptekar}, R.~L. and {Golenetskii}, S.~V. and {Pal'shin}, V.~D. and {Oleynik}, Ph. P. and {Tsvetkova}, A.~E. and {Ulanov}, M.~V. and {Cline}, T.~L. and {Hurley}, K.},
        title = "{The Second Konus-Wind Catalog of Short Gamma-Ray Bursts}",
      journal = {\apjs},
         year = 2016,
        month = may,
       volume = {224},
       number = {1},
          eid = {10},
        pages = {10},
          doi = {10.3847/0067-0049/224/1/10},
archivePrefix = {arXiv},
       eprint = {1603.06832},
 primaryClass = {astro-ph.HE},
       adsurl = {https://ui.adsabs.harvard.edu/abs/2016ApJS..224...10S}
}

@ARTICLE{Tsvetkova2017,
       author = {{Tsvetkova}, A. and {Frederiks}, D. and {Golenetskii}, S. and {Lysenko}, A. and {Oleynik}, P. and {Pal'shin}, V. and {Svinkin}, D. and {Ulanov}, M. and {Cline}, T. and {Hurley}, K. and {Aptekar}, R.},
        title = "{The Konus-Wind Catalog of Gamma-Ray Bursts with Known Redshifts. I. Bursts Detected in the Triggered Mode}",
      journal = {\apj},
         year = 2017,
        month = dec,
       volume = {850},
       number = {2},
          eid = {161},
        pages = {161},
          doi = {10.3847/1538-4357/aa96af},
archivePrefix = {arXiv},
       eprint = {1710.08746},
 primaryClass = {astro-ph.HE},
       adsurl = {https://ui.adsabs.harvard.edu/abs/2017ApJ...850..161T}
}

@ARTICLE{Tsvetkova2021,
       author = {{Tsvetkova}, Anastasia and {Frederiks}, Dmitry and {Svinkin}, Dmitry and {Aptekar}, Rafail and {Cline}, Thomas L. and {Golenetskii}, Sergei and {Hurley}, Kevin and {Lysenko}, Alexandra and {Ridnaia}, Anna and {Ulanov}, Mikhail},
        title = "{The Konus-Wind Catalog of Gamma-Ray Bursts with Known Redshifts. II. Waiting-Mode Bursts Simultaneously Detected by Swift/BAT}",
      journal = {\apj},
         year = 2021,
        month = feb,
       volume = {908},
       number = {1},
          eid = {83},
        pages = {83},
          doi = {10.3847/1538-4357/abd569},
archivePrefix = {arXiv},
       eprint = {2012.14849},
 primaryClass = {astro-ph.HE},
       adsurl = {https://ui.adsabs.harvard.edu/abs/2021ApJ...908...83T}
}

@ARTICLE{Woosley93,
       author = {{Woosley}, S.~E.},
        title = "{Gamma-Ray Bursts from Stellar Mass Accretion Disks around Black Holes}",
      journal = {\apj},
         year = 1993,
        month = mar,
       volume = {405},
        pages = {273},
          doi = {10.1086/172359},
       adsurl = {https://ui.adsabs.harvard.edu/abs/1993ApJ...405..273W}
}

\end{document}